\documentclass[aps,twocolumn,amsmath,amssymb,superscriptaddress,nofootinbib,prx]{revtex4-2}
\usepackage[utf8]{inputenc}
\usepackage{graphicx}
\usepackage{float} 
\usepackage[tight]{subfigure}
\usepackage{units} 
\usepackage{natbib} 
\usepackage{multirow}
\usepackage{cancel}
\usepackage{comment}

\usepackage[a4paper]{hyperref}
\hypersetup{colorlinks=true,
linktoc=all,
linkcolor=blue,
breaklinks=true,
citecolor=blue,
urlcolor=blue}

\usepackage[english]{babel}
\AtBeginDocument{%
    \newwrite\bibnotes
    \def\bibnotesext{Notes.bib}
    \immediate\openout\bibnotes=\jobname\bibnotesext
    \immediate\write\bibnotes{@CONTROL{REVTEX42Control}}
    \immediate\write\bibnotes{@CONTROL{%
    apsrev42Control,author="08",editor="1",pages="1",title="0",year="1"}}
     \if@filesw
     \immediate\write\@auxout{\string\citation{apsrev42Control}}%
    \fi
}%

\newcommand{\deriv}[2]{\frac{\mathrm{d} #1}{\mathrm{d} #2}}
\newcommand{\T}{\mathcal{T}}

\usepackage{array} 
\newcolumntype{C}[1]{>{\centering\arraybackslash}p{#1}}

\begin{document}

\title{Thermodynamic inference in molecular motors: a Martingale approach}

\author{Adrián Nadal-Rosa}
\affiliation{Institute for Cross-Disciplinary Physics and Complex Systems (IFISC), UIB-CSIC,
Campus Universitat de les Illes Balears E-07122, Palma de Mallorca, Spain.}
\affiliation{Theoretical and Computational Systems Biology Program,
Institute for Integrative Systems Biology (I2SysBio),
CSIC-UV, Catedrático Agustín Escardino Benlloch 9,
46980 Paterna, Spain}

\author{Gonzalo Manzano}
\affiliation{Institute for Cross-Disciplinary Physics and Complex Systems (IFISC), UIB-CSIC,
Campus Universitat de les Illes Balears E-07122, Palma de Mallorca, Spain.}

	
\begin{abstract}
    Molecular motors are in charge of almost every process in the life cycle of cells, such as protein synthesis, DNA replication, and cell locomotion, hence being of crucial importance for understanding the cellular dynamics. However, given their size scales on the order of nanometers, direct measurements are rather challenging, and the information that can be extracted from them is limited. In this work, we propose strategies based on martingale theory in stochastic thermodynamics to infer thermodynamic properties of molecular motors using a limited amount of available information. In particular, we use two recent theoretical results valid for systems arbitrary far of equilibrium: the integral fluctuation theorem (IFT) at stopping times, and a family of bounds to the maximal excursions of entropy production. The potential of these strategies is illustrated with a simple model for the F1-ATPase rotary molecular motor, where our approach is able to estimate several quantities determining the thermodynamics of the motor, such as the rotational work of the motor performed against an externally applied force, or the effective environmental temperature.
\end{abstract}

\maketitle

\section{Introduction} \label{sec:intro}

Molecular motors are a class of nanometer-sized macromolecular complexes capable of converting some source of energy, typically the free energy released by hydrolysis of ATP molecules, into mechanical work that is used to perform some biological function~\cite{book_MolCellBiology}. They are present in all living organisms and are responsible for a wide range of cell biology processes such as intracellular and extracellular transport (myosins~\cite{Spudich2001_MyosinSwing}), cell locomotion (rotary motors at the base of flagella~\cite{Minamino2008_MolMotorFlagela}), nucleic folding (DNA topoisomerases~\cite{Champoux2001_DNATopoisomerases}), replication (DNA polymerase~\cite{Steitz1999_DNAPolumerases}), transcription (RNA polymerase~\cite{ADarst2001_BacterialRNAPolymerase}), translation (ribosomes~\cite{Maguire2001_Ribosome, Ramakrishnan2002_Ribosome}) and for almost every cell life process.

Being nanoscale objects composed of soft material, molecular motors have energy scales at ambient temperature comparable to the size of thermal fluctuations $k_{\mathrm{B}} T$, where $k_{\mathrm{B}}$ is the Boltzmann constant~\cite{Sivak2020}. Therefore, their functioning is largely affected by environmental fluctuations that come from the frequent collision with molecules and proteins in the surrounding medium. Despite fluctuations, in order to perform their function, the motors experience a sequence of conformational changes in their structure that must, on average, make them evolve in a specific direction. For example, when ribosomes are translating an mRNA sequence, they need to translocate along the genetic chain in a specific direction and order, i.e. they need to sequentially read the chain from top to bottom. As a consequence, molecular motors need unbalanced transitions, making living systems truly out-of-equilibrium systems~\cite{Gnesotto2018_LivingSystems}. Nonequilibrium statistical mechanics thus arises as a natural framework in this context to unveil the physical principles underlying biological activity. In particular, stochastic thermodynamics~\cite{book_Peliti} has been widely used for their description (see e.g. the reviews in Refs.~\cite{Tawada2001, Astumian2010, Seifert_2012, Sivak2020, Speck2021_MolMach} and references therein). 

From a theoretical point of view, the dynamics of molecular motors is often modeled by means of--typically Markovian-- continuous-time stochastic processes. Examples include molecular motors that randomly jump between discrete states, corresponding to the different structural configurations that they adopt during their operation. This description is motivated by a time-scale separation between such metastable mesostates~\cite{Speck2021_MolMach}. In these cases, the operation of the molecular motor is performed by completing cycles in the state space, which generally require particular chemical reactions to take place, coinciding with specific orientations of the structure of the motor, and from the presence of chemical products of a previous reaction. 
In this context, despite the different chemical and mechanical components defining the motor states and their transitions, the only available information to track the stochastic evolution of the motor usually comes from a finite set of observable transitions involving a visible movement of the motor, such as mechanical steps or rotations in some of its units.

One of the most important molecular motors in which we focus in this work is the ATP synthase complex~\cite{book_MolCellBiology}, responsible for the synthesis of ATP, the energy currency of life. It appears in practically all living organisms, from prokaryotic cells to eukaryotic cells. The complex is composed of two main parts: the F0 domain, embedded in the cellular membrane, and the F1 domain, also called F1-ATPase. The latter is a rotary motor capable of using the energy released by translocation of protons across the F0 domain to synthesize or hydrolize ATP molecules~\cite{Itoh2004, Rondelez2005}. The F1-ATPase motor has been widely studied since the end of the 90s, both experimentally \cite{Noji1997_DirectObsF1ATPase, Yasuda1998_Rot120Steps} and theoretically \cite{Oster1998_EnergyTransductionF1ATPase}. The thermodynamic operation of the motor has been accurately explored in the last decade, assessing some of the main features of the motor and suggesting a nearly 100\% thermodynamic efficiency in its transduction of chemical energy to mechanical work~\cite{Toyabe2011,Toyabe2010_Energetics}. In this context, the so-called fluctuation theorem (FT)~\cite{Evans2002, Jarzynski2011} has been combined with single-molecule experiments to provide improved estimations of the rotatory torque in the motor~\cite{Hayashi2010_FluctTheoremF1ATPase}, which constitutes one of the first applications of universal relations in stochastic thermodynamics for inference purposes~\cite{Andrieux06,Hayasi2012,Ritort2012}. This inference concept~\cite{Seifert_2019} has also been applied and extended to many other active biological processes, and to other universal non-equilibrium relations, such as the relation between dissipation and irreversibility~\cite{Roldan10,Martinez19, Roldan21,Lynn21,Kapustin24,Harunari24} , the thermodynamic uncertainty relation~\cite{Barato2015,Pietzonka2016, Dechant2021, Puglisi2023}, the thermodynamic speed limits~\cite{Ito2020,Ito2020b} or the infimum law of martingale theory~\cite{Edgar2017_InfStopTimes,Neri22,Neri2023}. 

In this work we propose two novel (unexplored) inference strategies on molecular motors based on recent results in stochastic thermodynamics regarding the martingale theory for stochastic entropy production~\cite{Edgar2023_MartingalesPhysicists}. The first strategy is based on the (integral) fluctuation theorem at stopping times derived in Refs.~\cite{Edgar2017_InfStopTimes,Gonzalo2021_StopTGambDemons,Gonzalo2024_AbsIrrev}. A stopping time is the first time that a stochastic trajectory meets a specific predefined criterion, e.g. when some stochastic observable of the system takes a specific value or surpasses a given threshold. Typical examples indeed include first-passage times, or escape times, but also any combination of them is a valid stopping time. The second strategy instead uses universal bounds for the maximal excursions of entropy production and related currents derived in Ref.~\cite{GonzaloEdgar2022_ExtSt}, in the context of steady-state thermal machines. As we shall see, both strategies turn out to be useful for extracting information about the energetics of the motors and their parameters, as well as the (effective) temperature of the environment, as exemplified in the ATP synthase complex rotatory motor.

This paper is structured as follows. In Sec. \ref{sec:strategies} we briefly review recent results in stochastic thermodynamics using {  martingale} theory, and we introduce the two main strategies for thermodynamic inference in discrete-state molecular motors. In Sec. \ref{sec:model} we explain the model used to describe the F1-ATPase rotatory motor, and we present the results obtained using the proposed strategies for different thermodynamic quantities of interest in the model. We summarize our findings and discuss implications and possible extensions of our work in Sec. \ref{sec:conclusions}. Details on the F1-ATPase model dynamics, their parameters, and on the derivations and results are given in the appendices.

\section{Inference strategies using stopping times}\label{sec:strategies}

In the following, we review relevant theoretical results on martingale theory for stochastic thermodynamics and introduce the two main inference strategies we propose in a generic configuration for discrete-state molecular motors performing cycles in its state space.
{  For further details about the results of martingale theory in stochastic thermodynamic relevant to our strategies see Appendix \ref{ap:Martingale_Processes}}.

\subsection{IFT at stopping times} \label{sec:strategies_IFT}

A first version of the IFT for entropy production at stopping times in nonequilibrium steady-states was introduced in Ref.~\cite{Edgar2017_InfStopTimes} (see also Refs.~\cite{Chetrite18,Neri19}). The IFT was then extended in Ref.~\cite{Gonzalo2021_StopTGambDemons} for generic nonequilibrium dynamics, leading to the possibility of implementing successful gambling strategies that apparently violate the second law using stopping times:
\begin{equation}
    \left<e^{-S_{\mathrm{tot}}(\mathcal{T})-\delta(\mathcal{T})}\right>_{\mathcal{T}} = 
    1,
    \label{eq:IFT_stopping}
\end{equation}
where $S_{\mathrm{tot}}(\mathcal{T})$ stands for the (total) stochastic entropy production~\cite{Seifert_2012,Pigolotti17} in a single trajectory of the system $X_{[0,\mathcal{T}]} = \{ x(t): 0 \leq t \leq \mathcal{T}\}$ described by an stochastic observable of the system $x$ up to a generic (bounded) stopping time $\mathcal{T} \leq \tau$ with $\tau$ a limit time. The stochastic entropy production measures the irrevesibility of single trajectories as $S_{\mathrm{tot}}(\mathcal{T}) = \ln[\mathbb{P}(X_{[0,\T]})/\tilde{ \mathbb{P}}(\tilde{X}_{[0,\T]})]$ by comparing their path probabilities with the path probabilities of their time-reversed twins $\tilde{X}_{[0,\T]}$ in the time-reversed process. This reversed process begins at the final distribution of the forward one at the limit time $\tau$ and is let to evolve backwards (using the time-reversed driving protocol w.r.t. the forward process). Above $\delta_\tau(\mathcal{T})$ is the so-called stochastic distinguishability:
\begin{equation}
    \delta_\tau(\mathcal{T}) = \ln\frac{\rho_{x_{\mathcal{T}}}(\mathcal{T})}
    {\tilde{\rho}_{x_{\mathcal{T}}}(\tau - \mathcal{T})},
    \label{eq:delta}
\end{equation}
\noindent 
where $\rho_{x_{\mathcal{T}}}(\mathcal{T})$ is the probability of finding the system in state $x(\mathcal{T})$ in which the stopping condition is met at time $\mathcal{T} \leq \tau$, and $\tilde{\rho}_{X_{\mathcal{T}}}(\mathcal{T})$ is the probability of finding the system in the same state but in the time-reversed process run up to time $\tau -\mathcal{T}$. We notice that if the stopping condition is always met in the stationary state, then $\rho_x = \tilde{\rho}_x \equiv \pi_x$ is a time-independent distribution, and therefore $\delta_\tau(\mathcal{T}) = 0$, so that Eq.~\eqref{eq:IFT_stopping} reduces to the NESS IFT, $\langle e^{-S_\mathrm{tot}(\T)}\rangle_\T =1$, derived in Ref.~\cite{Edgar2017_InfStopTimes}. Here and in the following we denote the stationary distribution of the system states as $\pi_x$.

The IFT in Eq.~\eqref{eq:IFT_stopping}  has been experimentally tested in driven-dissipative systems under isothermal environmental conditions in both a nanoelectronic platform implementing a single electron box~\cite{Gonzalo2021_StopTGambDemons}, as well as for a Brownian particle trapped with optical tweezers~\cite{Albay23,Albay24}. There, the choice of smart stopping strategies allowed the detection of negative entropy production at stopping times in agreement with the generalized second law $\langle S_\mathrm{tot} \rangle_\T  \geq - \langle \delta_\tau \rangle_{\T}$, following directly from Eq.~\eqref{eq:IFT_stopping}.

Recently, the above fluctuation theorem was generalized in Ref.~\cite{Gonzalo2024_AbsIrrev} to setups that include absolute irreversibility~\cite{Murashita2014_AbsIrr}, where Eq.~\eqref{eq:IFT_stopping} may diverge. Most common situations leading to absolute irreversibility include setups with an initial distribution restricted in the state space, where trajectories $X_{[0,\tau]}$ with zero probability (as imposed by the initial condition) may have reversed twins $\tilde{X}_{[0,\tau]}$ with non-zero probability in the time-reversed process. The generalized IFT with absolute irreversibility reads:  
\begin{equation}
    \left<e^{-S_{\mathrm{tot}}(\mathcal{T})-\delta(\mathcal{T})}\right>_{\mathcal{T}} = 
    1-\Gamma,
    \label{eq:IFT_stopping_Irr}
\end{equation}
\noindent 
where $0 \leq \Gamma \leq 1$ is given by the expression:
\begin{equation}
    \Gamma = \int_0^{\tau} \mathrm{d}\mathcal{T} P(\T)\sum_{X_{[0,\mathcal{T}]} \in AI} \mathbb{\tilde{P}} 
    (\tilde{X}_{[\mathcal{T},0]} | \T)e^{-\delta_\tau(\mathcal{T})}.
    \label{eq:Gamma}
\end{equation}
\noindent Here $P(\T)$ is the probability of stopping at $\T$ and $\mathbb{\tilde{P}}(\tilde{X}_{[0,\T]}|\T)$ is the path probability of a reversed trajectory $\tilde{X}_{[0,\T]}$ that verified the stopping condition for the first time at $\T$.
The set $AI$ in the sum contains all trajectories that have zero probability in the forward process, $\mathbb{P}(X_{[0,\mathcal{T}]}) = 0$, but not in the backward one,
$\mathbb{\tilde{P}}(\tilde{X}_{[\mathcal{T},0]}) \neq 0$. We notice that $\Gamma$ is the average of 
$e^{-\delta_\tau(\mathcal{T})}$ over trajectories in the set $AI$. Moreover, when $\delta_\tau(\mathcal{T})=0$, $\Gamma$ becomes the 
total probability of stopped trajectories in set $AI$, that is, a measure of the total number of unpaired trajectories under the stopping protocol.

We are interested in using the universal relations in Eq.~\eqref{eq:IFT_stopping} and Eq.~\eqref{eq:IFT_stopping_Irr} to infer thermodynamical properties of molecular motors, related to the stochastic entropy production at stopping times $S_{\rm tot}(\T)$, which can be alternatively written in general as~\cite{Seifert_2019}:
\begin{equation}
    S_{\mathrm{tot}}(\T) = 
    \Delta S_{\mathrm{sys}}(\T) - \beta Q(X_{[0,\T]}).
    \label{eq:StotExplicit}
\end{equation}
\noindent 
Here $\Delta S_{\mathrm{sys}}(\T) = -\ln \rho_{x_\T}(\T) + \ln \rho_{x_0}(0) + \Delta S_\T^{\rm int}$ is the change of the system entropy (we use $k_{\mathrm{B}} = 1$ units), which, in general, includes both the change in the surprisal of the motor distribution and the change in the intrinsic entropy of their corresponding mesostates. For simplicity, in the following we focus on mesostates with approximately equal intrinsic entropy, i.e. $\Delta S_\T^{\rm int} \simeq 0$ and call them simply system states. The second term in Eq.~\eqref{eq:StotExplicit} stands for the entropy exchanged with the environment associated with the net heat $Q(X_{[0,\T]})$ that entered the system during the trajectory $X_{[0,\T]}$ at the environmental temperature $T = 1/\beta$. 

Molecular motors typically work under steady-state conditions. Moreover, in typical experimental situations, only few system states and transitions between them are actually observable. 
In such situations, none of the terms in Eq.~\eqref{eq:StotExplicit} may be fully known from the observable transitions. As we shall see in the following, this difficulty can be circumvented by choosing adequate stopping conditions that allow us to directly relate the entropy production with both the quantities of interest to be inferred, and the observable transitions during the stochastic trajectories of the system. 

\subsection{Counting cycles from observable transitions}

We propose stopping conditions that are able to select trajectories $X_{[0,\T]}$ that perform a prefixed number of cycles in state space in one or the opposite directions. 
For that purpose, we assume that each different cycle in the state space of the motor has at least one associated observable transition that can be used to signal the completion of the cycles. For the sake of simplicity, in the following we focus on the case of unicyclic phase space. (However, the same reasoning straightforwardly applies to multicyclic motors, each of which having at least one corresponding observable transition.) In this scenario, the number of cycles performed in a given direction up to some time $t$, can be determined by the (net) flux in the ``signaling" transition $N(t)$. Arbitrarily choosing one of the cycle directions as the ``forward" one, $N(t)$ is the number of times a jump in that transition is observed in the forward direction minus the number of times it is observed in the opposite direction. 

\begin{figure}[tb]
    \centering
    \includegraphics[width=\linewidth]{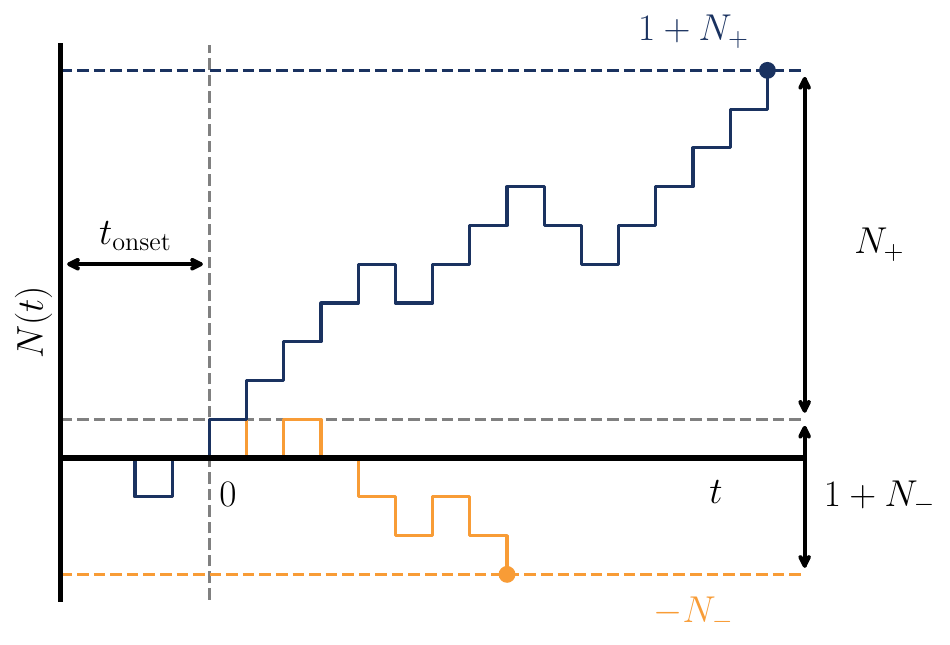}
    \caption{Illustration of the stopping protocol introduced in the main text with sample trajectories from the F1-ATPase rotatory motor (see Sec.~\ref{sec:model}). The current in the signalling transition $N(t)$ is monitored over time after the first transition is observed (after a time $t_{\rm onset}$), where we place the origin of times for the trajectories. Trajectories are stopped either when the current reaches the threshold value $N_+ +1$ (blue trajectory) performing $N_+$ cycles in the forward direction, or $-N_-$ (orange trajectory) corresponding to $N_-$ cycles in the backward direction after the starting point at $t_{\rm onset}$.}
    \label{fig:StoppingProtocol}
\end{figure}

The stopping condition we propose consists on a double threshold first-passage time for the current $N(t)$, as illustrated in Fig.~\ref{fig:StoppingProtocol}: the dynamics stops when the net number of signaling transitions performed by the motor reaches either the upper threshold $N_+ + 1$ or the lower threshold $-N_{-}$, with positive numbers $N_+ \neq N_{-}$ in general. We settle the starting point of the trajectories $X_{[0,\tau]}$ after the first signalling transition in the forward direction is detected, where we place the origin of time ($t_{\rm onset}$ in Fig.~\ref{fig:StoppingProtocol}).
We then let evolve the motor until the net number of jumps in the signalling transition reaches either the threshold $N_++1$ or $-N_-$, which corresponds to having trajectories with $N_+$ net transitions in the forward direction (i.e. from the starting point) or $N_- +1$ transitions in the backward direction, respectively. Moreover, here we will consider a limit time $\tau$ large enough to guarantee that the stochastic dynamics always verify the stopping condition before reaching the limit time, i.e. $\T < \tau$. This condition is ensured since the functional $e^{-S_\mathrm{tot}(\T) - \delta_\tau(\T)}$ remains bounded in the limit $\tau \rightarrow \infty$~\cite{Williams}, as guaranteed by the two-threshold first-crossing condition considered here.

 Let the motor states involved in the signalling transition be denoted as $x_1 \rightarrow x_2$. Therefore, the corresponding starting state of the motor for all trajectories is $x_2$ with probability one, i.e. the initial distribution at the starting point is a Kronecker delta, $\rho_{x}(0) = \delta_{x, x_2}$. Moreover, we assume that the final probability distribution of the motor when the stopping condition is met is always stationary (see also Sec. \ref{sec:model}), which is verified when the motor relaxation timescales are sufficiently fast compared to the time to reach the thresholds.

Since we assume a restricted initial condition in the state space of the motor, the relevant IFT that applies in this case is Eq.~\eqref{eq:IFT_stopping_Irr}, which contains the absolute irreversibility term $\Gamma > 0$. There we also have $\delta_\tau(\T) = 0$ as the stationary distribution is reached at the stopping time. We can then rewrite Eq.~\eqref{eq:IFT_stopping_Irr} by splitting the conditional average on the stopping conditions in terms of the thresholds as:
\begin{equation} \label{eq:FTsplit}
  \left<e^{-S_{\mathrm{tot}}(\mathcal{T})}\right>_{\mathcal{T}} = P_{-} e^{-S_\mathrm{tot}^-} + P_{+} e^{-S_\mathrm{tot}^+} = 1 - \Gamma,   
\end{equation}
with $P_+$ and $P_-$ the probabilities of first reaching the stopping thresholds $N_++1$ and $N_-$, respectively. The probability $P_+$ of reaching the threshold $N_++1$ (or $P_-= 1 - P_+$) can be obtained in experiments using a frequentist approach: After sampling a sufficiently large number of trajectories, $P_+$ is given by the number of trajectories stopped because they have reached the upper threshold over the total number of sampled trajectories. On the other hand, the absolute irreversibility term $\Gamma$ in Eq.~\eqref{eq:Gamma} becomes in this case the probability that time-reversed stopped trajectories do not reach $x_2$ at the origin of times. This quantity is cumbersome to calculate analytically, as it would require to obtain the path probabilities $\mathbb{{P}} (\tilde{X}_{[\mathcal{T},0]} | \T)$ for (reversed) trajectories that do not verify the stopping condition until a given time $\T$. However, we may approximate it as:
\begin{equation}
\begin{split}
\Gamma &= \int_0^\tau d\T P(\T)\sum_{X_{[0,\T]} \in AI} {\mathbb{P}}({  \tilde{X}_{[\T,0]}} | \T) \frac{\pi_{x_\T}}{\rho_{x_\T}(\T)}\\
&\simeq \int_0^\infty d\T P(\T)\sum_{x(0) \neq x_2}\pi_{x(0)} = 1-\pi_{x_2},  
\end{split}
\end{equation}
where $AI$ is the set of trajectories with $x(0) \neq x_2$, and we used the assumption that the steady state distribution has been reached at the limit time $\tau$. To obtain the second line, we approximated the transition probabilities inside $\mathbb{{P}} (\tilde{X}_{[\mathcal{T},0]} | \T)$ by the ones in $\mathbb{{P}} (\tilde{X}_{[\mathcal{T},0]})$, i.e. without conditioning on the stopping time $\T$. While this approximation may seem rough at first sight, we will later see that it is quite accurate in our example for the F1-ATPase rotatory motor in Sec.~\ref{sec:results_IFT_approximated}. 

We can next identify the entropy production at the stopping time for the two thresholds. The entropy production after $N_+$ net transitions in the forward direction are counted (threshold $N_++1$) is then:
\begin{equation}
    S_{\mathrm{tot}}^+ = -\ln\pi_{x_2} -  \beta N_+ Q_{\mathrm{cycle}},
    \label{eq:entropy_prod_forward}
\end{equation}
where $Q_{\rm cycle}$ stands for the heat transferred from the environment in a single cycle of the motor in the forward direction. Using the first law, the cycle heat can be related to the work sources acting on the motor, $Q_{\rm cycle}= W_{\rm cycle}^{\rm m} - W_{\rm cycle}^{\rm c}$, where $W_{\rm cycle}^{\rm m}$ stands for external mechanical work performed by the motor against external non-conservative forces acting during the cycle, and $W_{\rm cycle}^{\rm c} = \sum_{\alpha} \mu_\alpha \Delta N_\alpha$ is the net chemical work input in the motor, due to changes in the species concentration $\Delta N_\alpha$ with corresponding chemical potentials $\alpha$ along the cycle.

On the other hand, when the boundary $-N_{-}$ is reached, the entropy production becomes:
\begin{equation}
    S_{\mathrm{tot}}^- = -\ln\pi_{x_1} -  \beta q_{x_2 x_1} + \beta N_- Q_{\mathrm{cycle}},
    \label{eq:entropy_prod_backward}
\end{equation}
where $q_{x_2 x_1}$ is the heat absorbed in a single transition $x_2 \rightarrow x_1$. We remark that, in this case, it follows that $N_-$ cycles have been exactly produced when the threshold $N_{-}$ is reached, which means a net number of $N_{-} + 1$ transitions in the backward direction after the (onset) starting point. Moreover, the extra heat $q_{x_2 x_1}$ arise due to the fact that the final state ($x_1$) does not coincide with the initial one ($x_2$) in this case. 
That heat contribution can be generically rewritten as: 
\begin{equation} \label{eq:heat01}
q_{x_2 x_1} = \Delta E_{x_2 x_1} - \sum_\alpha \mu_\alpha \Delta N_{x_2 x_1}^{\alpha} + W_{x_2 x_1}^{\rm ext},    
\end{equation}
where $\Delta E_{x_2 x_1} = E_{x_1} - E_{x_2}$ is the energy difference between mesostates $x_1$ and $x_2$, $\Delta N_{x_2 x_1}^\alpha$ is the number of particles $\alpha$ exchanged with the reservoir in the transition, and {  $W_{x_2 x_1}^{\rm ext}$ is a contribution to the mechanical work performed by the system against an external force pointing in the $x_2 \rightarrow x_1$ direction}.

Plugging Eq.~\eqref{eq:entropy_prod_forward} and Eq.~\eqref{eq:entropy_prod_backward} into Eq.~\eqref{eq:FTsplit} and using that $P_+ + P_- = 1$, we obtain:
\begin{align} \label{eq:IFT_stopping_cycle_Irr}
     \left<e^{-S_{\mathrm{tot}}(\mathcal{T})}\right>_{\mathcal{T}} =&(1 - P_+) \pi_{x_1} e^{\beta\left[ q_{x_2 x_1} - N_-Q_{\mathrm{cycle}}\right]}  \nonumber \\ 
    &+ P_+ \pi_{x_2} e^{\beta N_+ Q_{\mathrm{cycle}}} = 1-\Gamma. 
\end{align}
The expression above following from the IFT~\eqref{eq:IFT_stopping_Irr} is a closed equation that relates the heat that enters the system in a cycle in state space $Q_{\mathrm{cycle}}$, with quantities depending only on the states $x_1$ and $x_2$, the probabilities $P_+$ and $\Gamma$, and the inverse temperature $\beta$. Eq.~\eqref{eq:IFT_stopping_cycle_Irr} is our first proposal for inferring thermodynamic quantities in molecular motors.

Remarkably, if the experimenter has access to the steady distribution values $\pi_{x_1}$ and $\pi_{x_2}$ (approximating $\Gamma \simeq 1 - \pi_{x_2}$) and $q_{x_1 x_2}$ is assumed to be known~\footnote{{  In particular, in equilibrium, the local detail balance of each pair of transitions may be used to asses the energy (or free energy) changes between mesostates~\cite{Seifert_2019}. Moreover, $\pi_{x_1}$ and $\pi_{x_2}$ might be obtained in the long-time limit as the fraction of time the system spends in each state, which may require extra auxiliary monitorable transitions from $x_1$ and $x_2$.}}, the heat per cycle $Q_\mathrm{cycle}$ can be inferred from Eq.~\eqref{eq:IFT_stopping_cycle_Irr}, which may be used to obtain two main quantities: 
First, if the external forces acting on the system are known, one can infer the chemical work performed by the motor, and hence the corresponding differences in chemical potentials. Second, if all the chemical potentials are already known (as we will mainly consider in the following), one could use this information to measure external forces in the nanoscale. 

On the other hand, if the experimenter has access to the heat exchanged during cycles $Q_{\rm cycle}$, to $q_{x_1 x_2}$ and to the stationary distribution values $\pi_{x_1}$ and $\pi_{x_2}$, then Eq.~\eqref{eq:IFT_stopping_cycle_Irr} can be used to infer the value of $\Gamma$, for which there are no alternatives in the literature. Finally, the method can also be used to infer the environmental temperature $T$ or quantities related to the heat on the transition $x_2 \rightarrow x_1$ [c.f. Eq.~\eqref{eq:heat01}] if the complementary quantities are known.

\subsection{Inference from IFT with approximations}

Nevertheless, in many situations of experimental interest only a few transitions among system states are observable. In that case, neither the stationary distribution of the states, $\pi_x$, nor the term $\Gamma$ would be accessible. To overcome this drawback, we propose a slightly different counting strategy and allow for some approximations. In this case, we assume to start the process directly from the stationary distribution $\pi_x$ in order to avoid absolute irreversibility. Therefore, now we start to count instances of the signalling transition without knowing exactly the initial state $x(0)$ (i.e. we do not wait any more for the first transition in the forward direction to happen to start counting). We then count the number of jumps in the signalling transition until the net flux $N(t)$ reaches one of the two thresholds $N_+$ or $-N_-$. 
We remark that in this scenario, both $\delta(\mathcal{T})$ and $\Gamma$ become zero. Besides, we approximate the entropy production at the two thresholds by:
\begin{equation}  \label{eq:Stot_aprox}
    \begin{split}
        S_{\mathrm{tot}}^{+} 
        &\simeq  - \beta N_{+} Q_{\mathrm{cycle}}, \\
        S_{\mathrm{tot}}^{-} 
        &\simeq  \beta N_{-} Q_{\mathrm{cycle}}.
    \end{split}
\end{equation}
The above approximation effectively neglects the entropy production during a (short) transient until the first signalling transition in either direction takes place. If the values $N_{\pm}$ are taken large enough, this transient is genuinely negligible, and the approximation becomes reliable. Eq.~\eqref{eq:FTsplit} then reads in this case:
\begin{equation}
(1 - P_+) e^{\beta N_- Q_{\mathrm{cycle}}} +   P_+ e^{-\beta N_+ Q_{\mathrm{cycle}}} \simeq 1,
    \label{eq:IFT_stopping_cycle_stst}
\end{equation}
which can be seen as an alternative proposal to Eq.~\eqref{eq:IFT_stopping_Irr}. The simplifications in Eq.~\eqref{eq:IFT_stopping_cycle_stst} allow us to directly relate $Q_{\rm cycle}$ with $P_+$ and $\beta$, thus providing a simple way to infer system parameters such as chemical potentials or non-conservative external forces acting on the system during each cycle. We notice that the above equation is similar to expressions derived in Ref.~\cite{Edgar2017_InfStopTimes} for the first passage probabilities of entropy production in stationary processes under two thresholds, with the difference that here the entropy production at the thresholds has been approximated by Eq.~\eqref{eq:Stot_aprox}.

In Sec.~\ref{sec:results_IFT_exact} we test the general validity of our strategy for a three-state model of the ATP-ase F1 rotatory motor. There, we show results corresponding to a hypothetical case in which stationary probabilities and $\Gamma$ are known, which we extract from simulations. Thereafter we compare with the parameters estimated using the approximate expression in Eq.~\eqref{eq:IFT_stopping_cycle_stst}, confirming the reliability of the approximations and the proposed strategy in a more practical situation, given typical experimental limitations. Finally, we also report results for the estimation of the temperature in the surrounding medium when other parameters of the motor are known.

\subsection{Bounds to maximal excursions of entropy production} \label{sec:strategies_bounds}

In Ref.~\cite{GonzaloEdgar2022_ExtSt}, universal families of bounds for the cumulative distribution of entropy production maxima and minima (and thus for the survival probability) within an interval $[0,\tau]$ in nonequilibrium steady-state processes were obtained:
\begin{equation}
\begin{gathered}
    \Pr[S_{\mathrm{max}}(\tau) \geq s_+] \leq e^{-p s_+} \left< e^{pS_{\mathrm{tot}}(\tau)} \right>, \\
    \Pr[S_{\mathrm{min}}(\tau) \leq -s_-] \leq e^{-p s_-} \left< e^{-pS_{\mathrm{tot}}(\tau)} \right>,
\end{gathered}
    \label{eq:bounds_SurvStot}
\end{equation}
\noindent where the short-hand notation $S_{\mathrm{max}}(\tau) = \max_{t\leq\tau} S_{\mathrm{tot}}(t)$ and $S_{\mathrm{min}}(\tau) = \min_{t\leq\tau} S_{\mathrm{tot}}(t)$, is used, $s_+, s_- > 0$ are a given threshold, and $p \geq 1$ is a real number leading to a one-parameter family of bounds. These bounds place restrictions on the probability that the maximum and minimum excursions of the entropy production in the interval $[0, \tau]$ overcome the thresholds $s_+$ and $-s_-$, respectively. Following the original work~\cite{GonzaloEdgar2022_ExtSt}, the probability bounds in Eq.~\eqref{eq:bounds_SurvStot} can be translated into optimal thresholds on the maximum and minimum entropy production that would not be surpassed with a given confidence probability during the interval. These bounds can also be used to bound the maxima and minima of the dissipated heat and the output work performed by steady-state heat engines and refrigerators. 

In the case of molecular motors, one may observe the output mechanical work performed by the motor by consuming some chemical fuel,i.e. from the chemical work that comes from breaking molecular bounds, such as those of ATP. The procedure shown in Ref.~\cite{GonzaloEdgar2022_ExtSt} can be directly generalized to bounds for the output mechanical work performed by a molecular motor,  as we show in Appendix~\ref{ap:BoundsWork}. As a result, we obtain 
optimal thresholds for the output mechanical work of the motor, which are guaranteed to not be overcome with a confidence probability $1 - \alpha$, with $\alpha>0$ typically small:
\begin{equation}
    w_\pm (t) = k_{\mathrm{B}}T \left(\frac{\eta}{1-\eta}\right) \ln\left[\underset{p\geq1}{\mathrm{min}}
    \frac{\left<e^{\pm p S_{\mathrm{tot}}(t)}\right>^{1/p}}{\alpha^{1/p}}\right],
    \label{eq:BoundsWork}
\end{equation}
\noindent where $\eta$ is the macroscopic efficiency of the motor, $\eta = -W_{\mathrm{cycle}}^{\rm m}/
W_{\mathrm{cycle}}^{\rm c} \leq 1$. 
This family of thresholds depends on the expectation values of $\left< e^{\pm pS_{\mathrm{tot}}(\tau)} \right>$, making it cumbersome to perform the minimization on $p$. 
Nonetheless, in the long-term limit $t\to\infty$, the minimization can be performed and the lower threshold in Eq.~\eqref{eq:BoundsWork} saturates to the following 
constant value (see Appendix~\ref{ap:BoundsWork} for details):
\begin{equation}
    w_- = -\frac{k_{\mathrm{B}}T\eta}{1-\eta}\ln\alpha.
    \label{eq:AssBoundWMinus}
\end{equation}

Assuming at least one visible transition in the state space that allows us to signal the mechanical work performed on a cycle from the stochastic current $N(t)$, that is, $N(t) W_{\rm cycle}^{\rm m}$, we 
relate the minima of the mechanical work output with the minima of $N(t)$. Using Eq.~(\ref{eq:AssBoundWMinus}) we then obtain an equation for the output mechanical work per cycle as:
\begin{equation}
    W_{\mathrm{cycle}}^{\rm m} = W_\mathrm{cycle}^{\rm c} + \frac{k_{\mathrm{B}}T\ln\alpha}{N_-^*},
    \label{eq:WmechInfLowBound}
\end{equation}
which represents our third main relation for thermodynamic inference. In the last equation, assuming the chemical work per cycle $W_{\mathrm{cycle}}^{\rm c}$ is known, the mechanical work $W_{\mathrm{cycle}}^{\rm m}$ per cycle can be inferred 
from the current values overcoming the minimum threshold, $N_{-}^*= w_{-}/W_{\rm cycle}^{\rm m}$ in interval $[0,\tau]$ for a given value of $\alpha$, and environmental temperature $T$. Analogously, if $W_{\rm cycle}^{\rm m}$ is known but $W_{\rm cycle}^{\rm c}$ is not, one can estimate the latter from the minimum current. 

To obtain $W_{\rm cycle}^{\rm m}$ or $W_{\rm cycle}^{\rm c}$ from experimental measurements, we propose the following steps: (1) run a sample of trajectories and count the number of signalling transitions in each direction to obtain the net current $N(t)$, (2) compute the minimum value of the current for each trajectory, $N_{\min}$, and construct a histogram for all trajectories sampled, (3) select threshold values $N_{-}^\ast$ in the left tail of the histogram and obtain the (cumulative) probability $\alpha = P[N_{\min} \leq N_{-}^\ast]$. 
The above procedure gives pairs $(N_{-}^\ast,
 \alpha)$ that can be plugged into Eq.~\eqref{eq:WmechInfLowBound} to obtain estimated values of the mechanical work per displacement, $W_\mathrm{cycle}^{\rm m}$, or the chemical input work, $W_\mathrm{cycle}^{\rm c}$. 

In Sec.~\ref{sec:results_bounds} we show the corresponding results of this second strategy when applied to the F1-ATPase rotatory motor model. Finally, we remark that Eq.~(\ref{eq:WmechInfLowBound}) can also be used to infer the temperature of the system, $T$, if the input chemical work per cycle, $W_{\mathrm{cycle}}^{\rm c}$, and mechanical work per cycle, $W_{\mathrm{cycle}}^{\rm m}$, are known.

\section{Illustration in the F1-ATP$\mathbf{ase}$ rotatory motor}\label{sec:model}

\begin{figure}
    \centering
    \includegraphics[width=\linewidth]{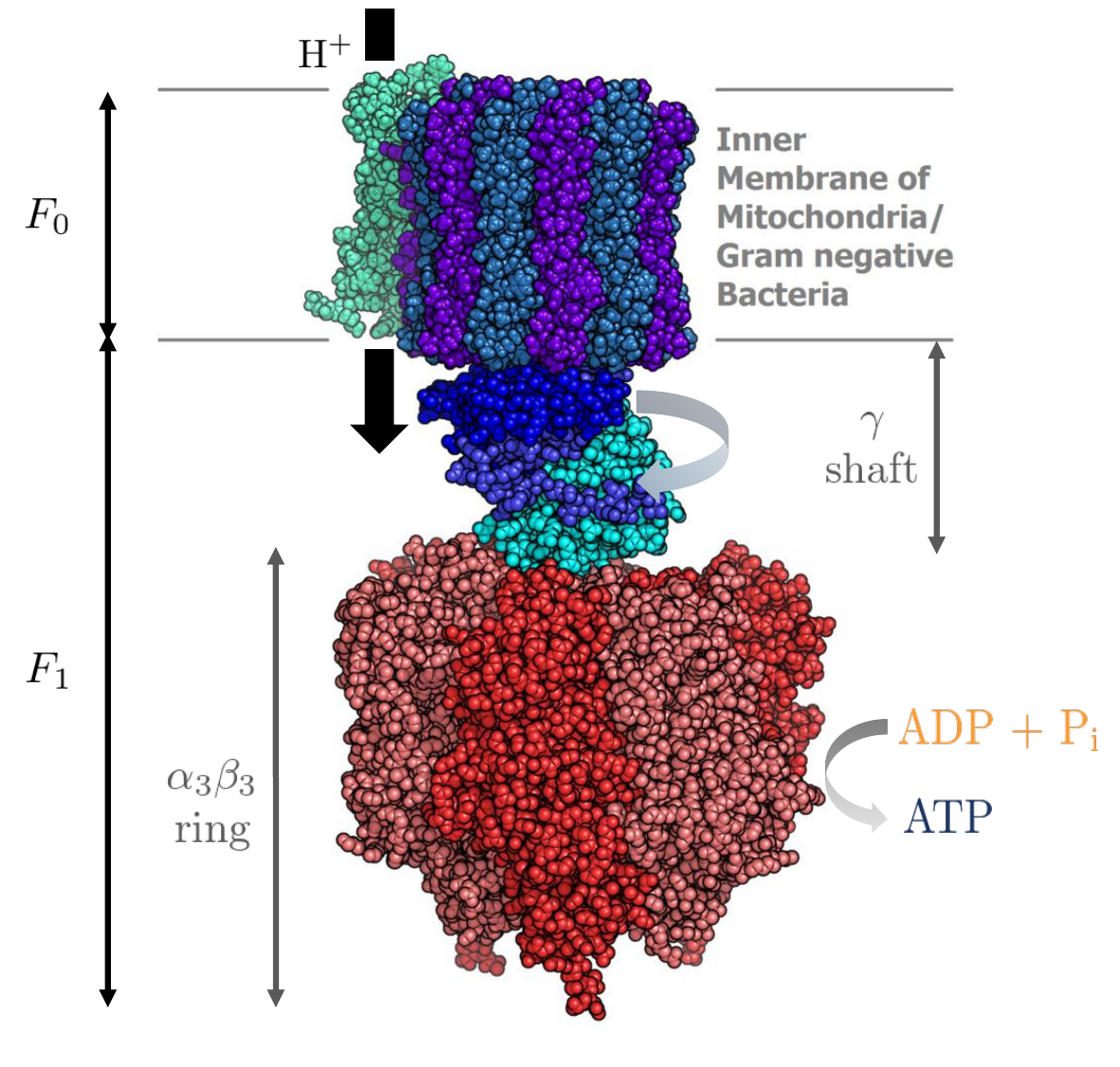}
    \caption{Scheme of the ATP synthase complex. The proton flux ($\mathrm{H}^+$) across the $F_0$ domain, embedded in the membrane, induces a rotation in the F1 domain, including the $\gamma$ shaft and the hexameric ring of $\alpha_3$ and $\beta_3$ proteins~\cite{figATP}.} 
    \label{fig:F1ATPase_MolecularScheme}
\end{figure}

We illustrate the methods proposed in the previous section for the experimentally relevant example of the F1-ATPase rotatory motor~\cite{Itoh2004,Rondelez2005,Noji1997_DirectObsF1ATPase, Yasuda1998_Rot120Steps}. The F1-ATPase is the F1 domain of the ATP synthase complex, the minimum unit of the complex that can act independently as a motor. It is mainly composed of a $\gamma$ shaft and a hexameric ring of $\alpha$ and $\beta$ proteins (see Fig.~\ref{fig:F1ATPase_MolecularScheme}). When working as part of the ATP synthase, the motor rotates due to the proton flux that enters the cell across the F0 domain. It makes the $\gamma$ shaft rotate, and hence the protein ring, which experiences conformational changes that allow the system to take $\mathrm{ADP}$ and inorganic phosphate $\mathrm{Pi}$, from the environment and synthesize $\mathrm{ATP}$. If the motor rotates in the reverse direction, it uses the energy stored in the bounds of the ATP molecules to pump protons outside the cell.

\begin{figure*}[t]
    \centering
    \includegraphics[width=0.9\textwidth]{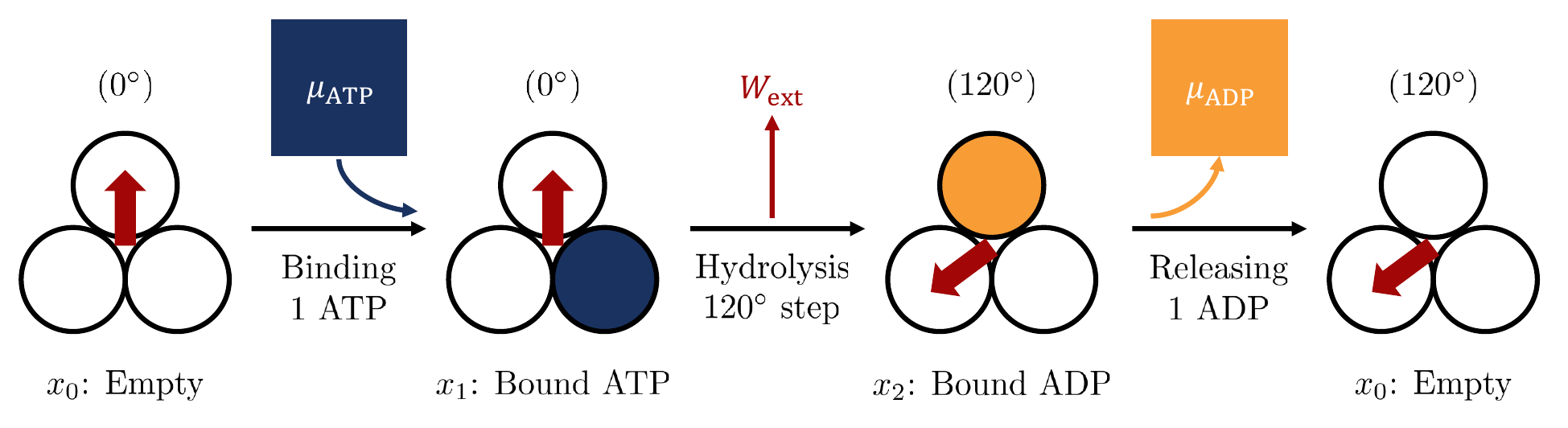}
    \caption{Three-states model for the F1-ATPase motor performing a cycle on the hydrolisis (anticlockwise direction) in three steps: Binding ATP ($x_0 \rightarrow x_1$), hydrolising ATP into ADP + P, accompanied by a $120^\circ$ rotation against an external torque ($x_1 \rightarrow x_2$), and releasing $ADP$ ($x_2 \rightarrow x_0$). Only transitions that complete the cycle are explicit for the sake of readability, but the reverse transitions are also possible. The system and the reservoirs of ATP and ADP (colored squares) are all embedded in a dissolution at a particular temperature, $T$.}
    \label{fig:Modelo}
\end{figure*}

We model the dynamics of the F1-ATPase as a continuous-time Markovian process over a discrete set of three states $\{x_0,~x_1,~x_2 \}$. 
These three states are depicted in Fig.~\ref{fig:Modelo}, and correspond to situations in which the motor is empty ($x_0$), when it bounds ATP ($x_1$), and when it bounds ADP ($x_2$). The transition responsible for motor movement is only $x_1 \leftrightarrow x_2$, associated with the hydrolysis ($x_1\to x_2$) and the synthesis ($x_2\to x_1$) of ATP. They involve a $\Delta \theta = 2\pi / 3 ~(120^{\circ})$ rotation during hydrolysis, and a $\Delta \theta = - 2 \pi/ 3$ rotation during synthesis. Here and in the following, we arbitrarily choose the direction of hydrolysis  as the ``forward" one, corresponding to an anticlockwise roation in Fig.~\ref{fig:Modelo}.

We consider the system embedded in a dissolution that acts as a common thermal bath at temperature $T$ for the three pairs of transitions. As depicted in Fig.~\ref{fig:Modelo},
transitions involving particle exchanges are mediated by a particle reservoir, with ATP or ADP molecules, in which only a single molecule is exchanged per transition.
We denote by $\mu_{\mathrm{ATP}}$ and $\mu_{\mathrm{ADP}}$ the chemical potentials of ATP and ADP molecules, with $\Delta \mu = \mu_{\mathrm{ATP}} - \mu_{\mathrm{ADP}} > 0$. Moreover, motivated by the experiments in~\cite{Hayashi2010_FluctTheoremF1ATPase}, we include an external mechanical torque $\tau_{\rm ext}$ acting against the transition $x_1 \rightarrow x_2$ (between the bounding of ATP and the bounding of ADP states). This leads to mechanical output work performed by the motor against the external force $W_{\mathrm{ext}} = \tau_{\rm ext} \Delta \theta$ along with the anticlockwise rotation of the motor ($\Delta \theta = 2 \pi /3$), which becomes negative (work input) in the clockwise direction $x_2 \rightarrow x_1$ ($\Delta \theta = -2 \pi /3$). In fact, the external torque can be used to bias the motor to rotate on average in the direction of ATP synthesis when $\tau_{\mathrm{ext}} > \Delta \mu/|\Delta \theta|$, at the price of the external input work.

The model above takes into account the main features that were experimentally observed in the 90s~\cite{Noji1997_DirectObsF1ATPase, Yasuda1998_Rot120Steps, Oster1998_EnergyTransductionF1ATPase}, and in the more recent Refs.~\cite{Toyabe2011,Toyabe2010_Energetics}, which stated that the F1-ATPase rotates with discrete $120^{\circ}$ steps with close to 100\% efficiency. 
More details on the motor rotational mechanism have recently been revealed using temperature and time-resolved cryo-electron microscopy~ \cite{Meghna2021_F1ATPaseCycle,Sobti2023,Murata2023}. For instance, the 120 degrees steps can be unfolded in a combination of several substeps coupled to different chemical reactions taking place in the hexameric ring~\cite{Meghna2021_F1ATPaseCycle}. However, in order to keep our analysis simple, we do not consider these details in the model.

The time evolution of the probability of finding the system at a given time $t$, in a particular state $x$, 
is thus given by the following master equation:
\begin{equation}
    \deriv{\rho_x(t)}{t} = \sum_{x'\neq x}[K_{x'x}\rho_{x'}(t) - K_{x x'}\rho_x(t)],
    \label{eq:master}
\end{equation}
where $K_{x x'}$ is the transition rate from state $x$ to state $x'$. The rates are consistent with the local detailed balance (LDB) condition (see e.g. Ref.~\cite{Seifert_2019}) and read:
\begin{equation}
    \begin{aligned}
        K_{x_0 x_1} &= k_{\mathrm{ATP}}~e^{-\frac{\beta}{2} \left(\Delta F_{x_0 x_1} - \mu_{\mathrm{ATP}}\right)}, \\
        K_{x_1 x_0} &= k_{\mathrm{ATP}}~e^{-\frac{\beta}{2} \left(\Delta F_{x_1 x_0} + \mu_{\mathrm{ATP}}\right)}, \\
        K_{x_1 x_2} &= k_{HS}~e^{-\frac{\beta}{2} \left(\Delta F_{x_1 x_2} + W_{\mathrm{ext}}\right)}, \\
        K_{x_2 x_1} &= k_{HS}~e^{-\frac{\beta}{2} \left(\Delta F_{x_2 x_1} + W_{\mathrm{ext}}\right)}, \\
        K_{x_2 x_0} &= k_{\mathrm{ADP}}~e^{-\frac{\beta}{2} \left(\Delta F_{x_2 x_0} + \mu_{\mathrm{ADP}}\right)}, \\
        K_{x_0 x_2} &= k_{\mathrm{ADP}}~e^{-\frac{\beta}{2} \left(\Delta F_{x_0 x_2} - \mu_{\mathrm{ADP}}\right)},
    \end{aligned}
    \label{eq:rates}
\end{equation}
where $\Delta F_{x_i x_j} = F_{x_j} - F_{x_i}$ are the changes in the equilibrium free energies of the system states, and $k_{\mathrm{ATP}}$, $k_{\mathrm{ADP}}$, and $k_{HS}$ are constants setting the timescales in each transition (bounding ATP, bounding ADP, and hydrolysis/ 
synthesis reaction, respectively). We further assume $\Delta F_{x_i x_j} = \Delta E_{x_i x_j}$, in agreement with the assumption of approximately equal intrinsic entropies of the states, $\Delta S_\T^{\rm int} \simeq 0$.
 
We perform comprehensive numerical simulations that are used to test the reliability of the different inference strategies in Sec.~\ref{sec:strategies}. Experimental measurements
are simulated as stochastic trajectories given by the Gillespie algorithm~ \cite{book_NumericalMethods}. The two inference targets are the external work introduced by the work reservoir per transition, $W_{\mathrm{ext}}$ and the temperature $T$ of the surrounding environment.

The values for all parameters used in the simulations are given in Tab.~\ref{tab:exp_values}. Their experimental validation is explained in Appendix~\ref{ap:ExpValues}. We note that with these parameters the time needed to reach the stationary state is, on average, an order of magnitude shorter than the time needed to perform a single rotation in the hydrolysis direction (approximately $0.1~\text{s}$), as shown in Appendix~\ref{ap:MasterEq}. This implies that the motor is always rotating in the stationary state, which guarantees the applicability of our approach.

\begin{table}[t]
    \centering
    \renewcommand{\arraystretch}{1.5} 
    \begin{tabular}{|C{0.9\linewidth}|} 
    \hline
    \textbf{Internal Energies} $\left[k_{\mathrm{B}} T ~\mathrm{units} \right]$ \\ \hline
    $\Delta E_{x_0 x_1} = 20$ (adapted from \cite{Oster1998_EnergyTransductionF1ATPase,Toyabe2010_Energetics}) \\ 
    $\Delta E_{x_1 x_2} = -17.5$ \cite{Oster1998_EnergyTransductionF1ATPase,Toyabe2010_Energetics} \\ 
    $\Delta E_{x_2 x_0} = -2.5$ (adapted from \cite{Oster1998_EnergyTransductionF1ATPase,Toyabe2010_Energetics}) \\ \hline 
    \textbf{Chemical potentials} $\left[k_{\mathrm{B}} T ~\mathrm{units} \right]$ \\ \hline
    $\mu_{\mathrm{ADP}} = 0$ (energy origin) \\
    $\mu_{\mathrm{ATP}} = 18$ (obtained from \cite{Toyabe2010_Energetics}) \\ \hline
    \textbf{Timescales} \\ \hline
    $k_{\mathrm{ATP}} = k_{\mathrm{ADP}} = k_{HS} = 165~\mathrm{Hz}$ (obtained from \cite{Noji1997_DirectObsF1ATPase}) \\ \hline
    \end{tabular}
    \caption{Summary of the experimental values employed to characterize the rates of the system. Approximated concentrations taken from different experimental setups for these working parameters are $[\mathrm{ATP}] = [\mathrm{ADP}] = 0.4~\mathrm{nM}$  and $[\mathrm{Pi}] = 1~\mathrm{mM}$, and temperature $T=298~\rm{K} ~(25^{\circ}\rm{C})$.
    }
    \label{tab:exp_values}
\end{table}

\subsection{Inference results from the exact IFT at stopping times} \label{sec:results_IFT_exact}

For F1-ATPase, we choose as a signaling transition the hydrolysis-synthesis transition ($x_1 \leftrightarrow x_2$), which is the only transition that involves a macroscopically observable angular displacement of the motor $\Delta \theta = \pm 2 \pi / 3$ in one or the opposite directions. Following the stopping protocol introduced in Sec.~\ref{sec:strategies_IFT}, we are interested in the net number of hydrolysis and synthesis transitions during a time interval, which can be obtained by recording the time evolution of the angular position of the motor. An illustration of the inference stopping protocol is shown in Fig.~\ref{fig:StoppingProtocol}. After observing the first rotation in the hydrolysis direction at $t_{\rm onset}$, we track the angular position of the motor until a net number of hydrolysis $N_+$ or synthesis transitions $N_-$, that is, when the angular position of the motor reaches thresholds $(N_+ + 1)|\Delta\theta|$ or $-N_-|\Delta\theta|$, respectively.
The heat that enters the system in a cycle in the hydrolysis direction is in this case:
\begin{equation}
    Q_{\mathrm{cycle}} = q_{x_0 x_1} + q_{x_1 x_2} + q_{x_2 x_0} = W_{\mathrm{ext}}-\Delta\mu,
\end{equation}
where the total output mechanical work in the cycle $W_{\mathrm{clycle}}^{\rm m} = W_{\mathrm{ext}} = \tau_{\rm ext} \Delta \theta$ is employed in the rotation $\Delta \theta$ against the external torque $\tau_{\rm ext}$, and the input chemical work $W_{\rm ext}^{\rm c} = \Delta\mu$ is obtained from the hydrolysis of ATP. Particularizing Eq.~\eqref{eq:IFT_stopping_cycle_Irr} to this case we obtain:
\begin{align} \label{eq:IFT_F1ATPase_Gamma}
     &\langle e^{-S_\mathrm{tot}(\T)}\rangle_\T = P_+ \pi_{x_2} e^{- \beta N_+( \Delta\mu - W_{\mathrm{ext}})}   \\ 
     &+ (1 - P_+) \pi_{x_1} e^{\beta(\Delta E_{21}-\Delta\mu)}
    e^{\beta N_-(\Delta\mu-W_{\mathrm{ext}})} = 1 - \Gamma, \nonumber
\end{align}
where the indexes $+$ and $-$ represent hydrolysis and synthesis, respectively.

\begin{figure}[tb]
    \centering
        \includegraphics[width=\linewidth]{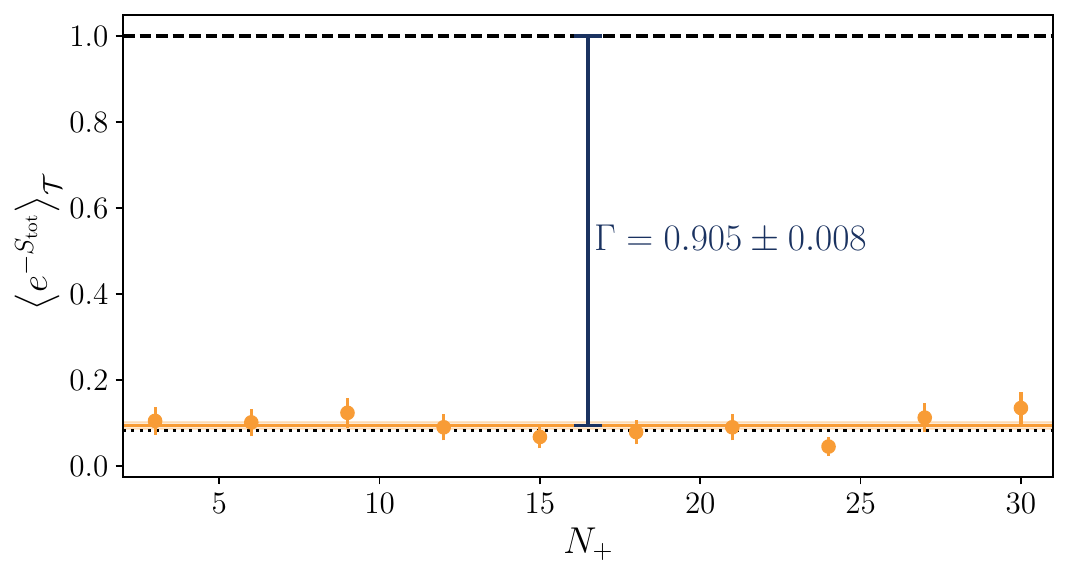}
    \caption{Inference of the absolute irreversibility term $\Gamma$ using the expression for IFT at stopping times in Eq.~\eqref{eq:IFT_stopping_cycle_Irr}. The functional $\langle e^{-S_{\rm tot}(\T)} \rangle_\T$ is obtained from sample trajectories under the stopping condition for thesholds $N_{-}=3$ and varying $N_+$ (yellow circles). Each value is obtained using $10^2$ trajectories. $\Gamma$ is estimated using the average of all different values. The external torque has been taken to be $\tau_{\rm ext} = 17~ k_B T / \Delta \theta$, and a surrounding temperature $T=298~{\rm K}$. Other parameters are as in Table~\ref{tab:exp_values}.}
    \label{fig:Results_WextInf_IFT_Gamma1}
\end{figure}

\begin{figure}
    \centering
        \includegraphics[width=\linewidth]{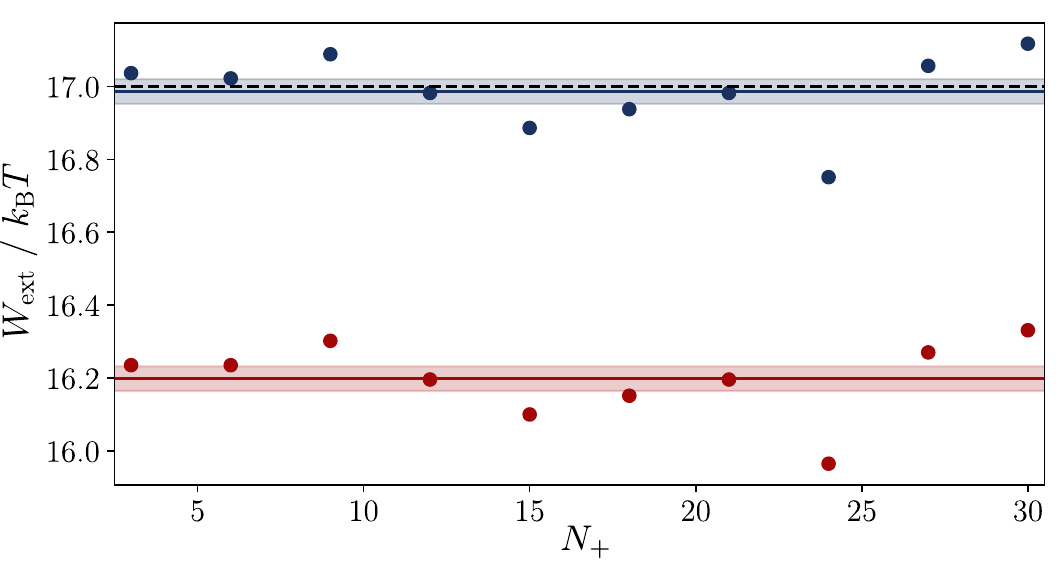}
    \caption{Inference of the output work in a rotation $W_{\mathrm{ext}}$ using the expression for the complete IFT at stopping times in Eq.~\eqref{eq:IFT_F1ATPase_Gamma} with $\Gamma$ (upper blue dots) and its inequality version without it (lower red dots). Again we set $N_{-} = 3$ and obtain estimations for different values of $N_+$ using $10^2$ trajectories for each value. Average estimated values are plotted as solid lines, $\bar{W}_{\mathrm{ext}} = 16.99 \pm 0.03 ~ k_{\mathrm{B}}T$ (blue) and $\bar{W}_{\mathrm{ext}} = 16.20 \pm 0.03 ~ k_{\mathrm{B}}T$ (red) with standardized errors given by the corresponding shaded region, which are compared with the true value of $W_{\rm ext}= 17~k_{\mathrm{B}} T$ used in the simulations (black dashed line) with $T = 298~{\rm K}$. All other parameters employed in the simulation are given in Tab.~\ref{tab:exp_values}.}
    \label{fig:Results_WextInf_IFT_Gamma2}
\end{figure}

In the following, we present our results for thermodynamic inference using the complete IFT in Eq.~\eqref{eq:IFT_F1ATPase_Gamma} with absolute irreversibility. As a consistency test, we first use the method to estimate the term $\Gamma$ (Fig.~\ref{fig:Results_WextInf_IFT_Gamma1}) from the probabilities $P_+$ (or $P_-$) obtained from the numerically sampled trajectories and compare with the approximate value $\Gamma \simeq 1 - \pi_{x_2}$. Then we use the method to estimate the external torque applied to the motor through the work $W_{\mathrm{ext}}$ (Fig. ~\ref{fig:Results_WextInf_IFT_Gamma2}) from the sampled trajectory probabilities $P_+$ (or $P_-$), and assuming that the complementary quantities are known. 

In Fig.~\ref{fig:Results_WextInf_IFT_Gamma1} the yellow points represent the estimation of the functional $\langle e^{-S_\mathrm{tot}(\T)} \rangle_\T$ from $10^2$ sample trajectories with a fixed lower threshold with $N_- = 3$ and for varying $N_+$. $\Gamma$ is obtained from the difference $1 - \langle e^{-S_\mathrm{tot}(\T)}\rangle_\T$, leading to an average value of $\bar{\Gamma} = 0.905 \pm 0.008$, compatible within two standard deviations with the approximate value $\Gamma \simeq 1 - \pi_{x_2} = 0.917$. The error bars are calculated from the standard deviation of the trajectories values and the final error by combining individual errors for each $N_+$ value that amounts to a total number of $10^3$ trajectories. For this case, we considered access to all the quantities appearing in the l.h.s. of Eq.~\ref{fig:Results_WextInf_IFT_Gamma1}, i.e. the probabilities $\pi_{x_1}$ and $\pi_{x_2}$, together with parameters $\Delta \mu$, $W_{\rm ext}$ and $\Delta E_{x_2 x_1}$.

In Fig.~\ref{fig:Results_WextInf_IFT_Gamma2} we instead present our results for the estimation of the output work $W_{\rm ext}$ in a $120^\circ$ rotation, which can be directly employed to obtain the external torque $\tau_{\rm ext} = W_{\rm ext}/(2 \pi/3)$. Two scenarios are considered, each with different available information about the system. In the first scenario, we assume knowledge of both the stationary probabilities $\pi_{x}$ and the absolute irreversibility term $\Gamma$, leading to inferred values of $W_{\mathrm{ext}}$ (blue dots) with mean $\bar{W}_{\mathrm{ext}} = 16.99 \pm 0.03 ~ k_{\mathrm{B}}T$ (blue dots), compatible within the error bars (shaded area) with the real values used in the simulations $W_{\mathrm{ext}} = 17~k_{\mathrm{B}} T$ for $T=298 {\rm K}$ [for other parameters, see Tab.~\ref{tab:exp_values}]. We then conclude that the method is reliable and precise for inferring the external torque applied, $\tau_{\rm ext} \simeq 8.12~k_{\mathrm{B}}T/\mathrm{rad}$. In the second scenario, we instead consider that the absolute irreversibility term $\Gamma$ is either unknown or not taken into account. In this case, only the bound $\langle e^{-S_{\mathrm{tot}}(\T)}\rangle_\T < 1$ can be used to obtain inferred values of $W_{\rm ext}$ (red points). This gives an estimated value of $\bar{W}_{\mathrm{ext}} = 16.20 \pm 0.03 ~ k_{\mathrm{B}}T$. We notice that this result is incompatible with the true value ($W_{\rm ext}= 17~k_{\mathrm{B}} T$), which highlights the importance of the absolute irreversibility term $\Gamma$ to obtain a precise estimate. However, the inferred value shows a relative deviation from the true one which is below $5~\%$, which is not a bad estimation after all. 
The statistical errors in both cases are again calculated as the standard deviation of the values obtained from the $10^2$ sampled trajectories for each $N_+$ value, leading to a total amount of $10^3$ trajectories. 

\subsection{Inference results from the IFT at stopping times with approximations} \label{sec:results_IFT_approximated}

We now test the strategy for inferring the external torque with the approximate expression derived in Sec.~\ref{sec:strategies_IFT}. The general Eq.~\eqref{eq:IFT_stopping_cycle_stst} reads for the F1-ATPase case:
\begin{equation}
    e^{-N_+\beta( \Delta\mu - W_{\mathrm{ext}})} P_+ + 
    e^{N_-\beta(\Delta\mu-W_{\mathrm{ext}})}(1- P_+) \simeq 1,
    \label{eq:IFT_F1ATPase_StSt}
\end{equation}
which allows for the direct estimation of the rotational work $W_\mathrm{ext}$ (and hence of the external torque $\tau_{\rm ext}$) from the probability $P_+$ (or $P_-$) and parameters $\Delta \mu$ and $\beta$ that do not require knowledge about the transitions among all three states of the model, unlike in the previous case.

\begin{figure}[t]
    \centering
    \includegraphics[width=\linewidth]{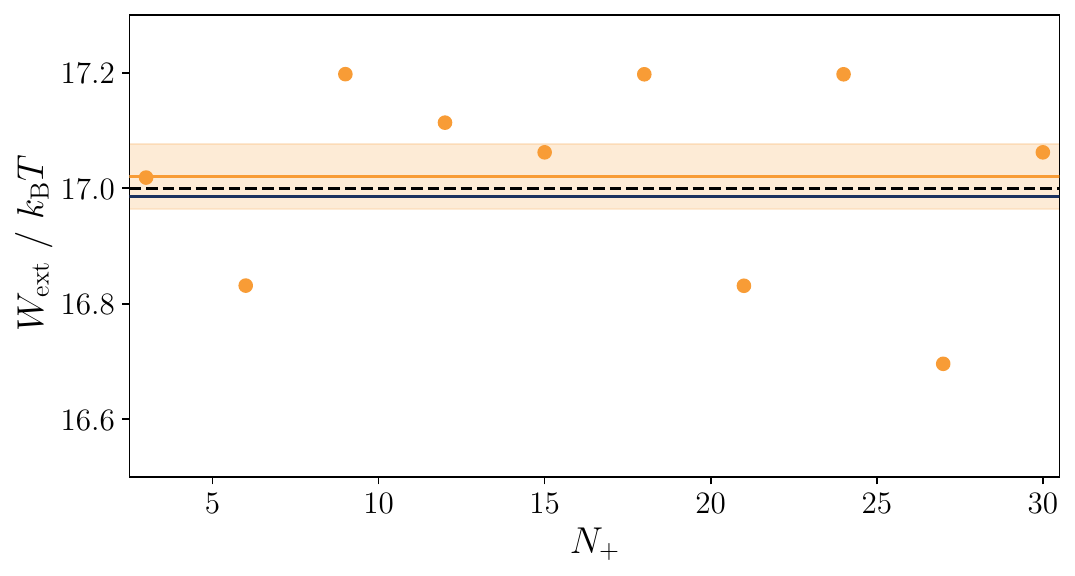}
    \caption{Inference of the output work in a rotation $ W_{\mathrm{ext}}$ from the approximate expression for the IFT at stopping times, Eq.~\eqref{eq:IFT_F1ATPase_StSt}. The estimations for different values of $N_+$ employ $10^2$ trajectories (yellow circles) and again $N_{-} = 3$ is fixed. The average estimated value (yellow solid line) is $\bar{W}_{\mathrm{ext}} = 17.02 \pm 0.06 ~ k_{\mathrm{B}}T$ with standardized errors (shaded region) is compared with the true value $W_{\rm ext}= 17 ~ k_{\mathrm{B}}T$ (black dashed line) and the result obtained from the complete IFT with absolute irreversibility $W_{\rm ext}= 16.99 ~ k_{\mathrm{B}}T$ (blue solid line). Here $T = 298 ~ {\rm K}$ and all other parameters are as in Tab.~\ref{tab:exp_values}.}
    \label{fig:Results_WextInf_IFT_StSt}
\end{figure}

The results for $W_{\mathrm{ext}}$ using Eq.~\eqref{eq:IFT_F1ATPase_StSt} are shown in Fig.~\ref{fig:Results_WextInf_IFT_StSt}, where again we use a fixed lower threshold $N_-=3$ and 10 values of $N_+$, as in the previous figures (yellow circles). The approximations for the stationary regime give an inferred value of $\bar{W}_{\mathrm{ext}} = 17.02 \pm 0.06 ~ k_{\mathrm{B}}T$ (solid yellow line), which is compatible within its error bars with the one used in simulations $W_{\rm ext} = 17 k_B T$ (dashed line). For the sake of comparison, we also plot the estimated value obtained from the complete IFT with absolute irreversibility in Eq.~\eqref{eq:IFT_F1ATPase_Gamma} obtained above (blue solid line); see Fig.~\ref{fig:Results_WextInf_IFT_Gamma2}. Surprisingly, despite the approximations performed in Eq.~\eqref{eq:IFT_F1ATPase_StSt}, this method provides a precise and accurate result comparable to that obtained with complete information. This makes us highlight it as the best strategy to follow, given the simplicity of the expressions and the minimal number of parameters needed in the estimation (which do not require knowledge of the stationary densities $\pi_x$).

\begin{figure}[t]
    \centering
    \includegraphics[width=\linewidth]{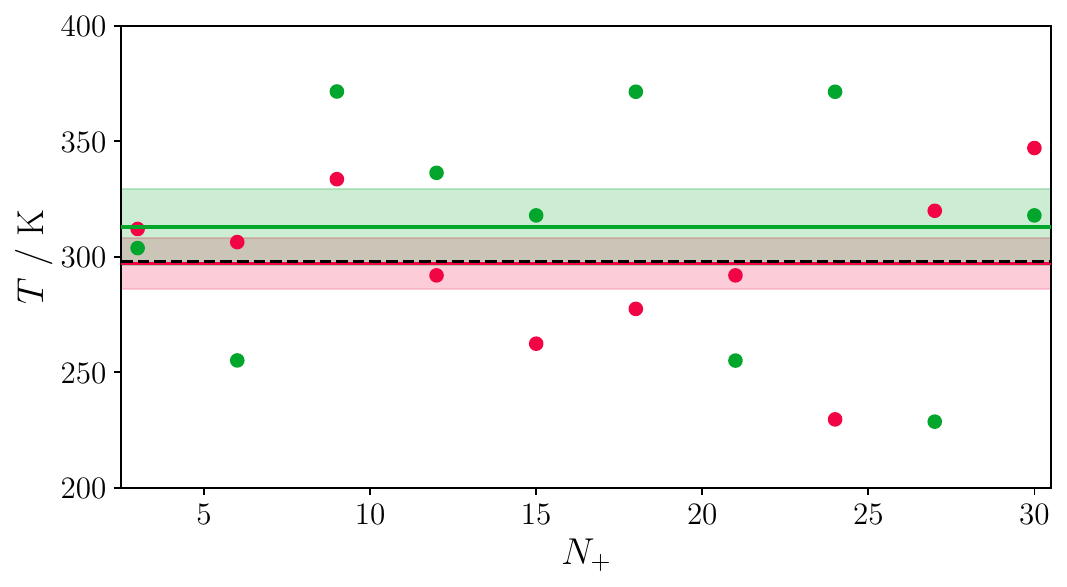}
    \caption{Inference of the effective temperature of the environment, $T$, from the approximate expression for the IFT at stopping times, Eq.~\eqref{eq:IFT_F1ATPase_StSt}. The estimations for different values of $N_+$ employ $10^2$ trajectories (green circles) and again $N_{-} = 3$ is fixed. The average estimated value (green solid line) is $\bar{T} = 313 \pm 17 ~ \mathrm{K}$ with standardized errors (shaded region). We compare with the result obtained from the complete IFT with absolute irreversibility \eqref{eq:IFT_F1ATPase_Gamma}, leading to $\bar{T} = 297\pm11 ~ \mathrm{K}$ (pink solid line). The true value of the temperature is $T= 298 ~ \mathrm{K}$ (black dashed line). All other parameters are as in Tab.~\ref{tab:exp_values}.}
    \label{fig:Results_TempInf_IFT_StSt}

\end{figure}

Finally, in Fig.~\ref{fig:Results_TempInf_IFT_StSt}, we show our results using this method to infer the effective temperature of the surrounding medium $T$ when the applied torque $\tau_{\rm ext}$ is known. We compare both inference strategies using the complete IFT with absolute irreversibility in Eq.~\eqref{eq:IFT_F1ATPase_Gamma} and the approximate expression in Eq.~\eqref{eq:IFT_F1ATPase_StSt} for steady-state dynamics. 
In the first case, the method infers a value of $T = 297 \pm 11 ~ \mathrm{K}$, compatible with the value used in the simulations, $T = 298 ~\mathrm{K}$ (dashed line) [see also Tab. \ref{tab:exp_values}]. The approximated IFT method gives instead a value of $T = 313 \pm 17 ~ \mathrm{K}$, which is also compatible with the true value. Our results hence indicate that both methods are also reliable and precise for the inference of the temperature.

It is worth remarking that in all the above tests, we used a small number of sampled trajectories ($10^2$) for the inferred values on purpose. The aim of this work is to provide reliable and feasible thermodynamic inference strategies. Thus, obtaining accurate results cannot rely on sampling a potentially unfeasible large number of trajectories for experimentalists. In this context, it is important to keep in mind the time that these measurements could eventually imply in a realistic experimental setup. In the F1-ATPase case, the motor completes a cycle in the forward (anticlockwise) hydrolysis direction at a time of the order of $0.1~\mathrm{s}$. The most time-consuming situation we considered corresponds to an upper threshold with $N_+ = 30$, which requires, on average, $3~\mathrm{s}$ to complete the stopping protocol. Recording the motion of $10^2$ trajectories would then take $300~\mathrm{s} = 5~\mathrm{min}$. That means that the proposed inferred method would be feasible under experimental conditions, since measurements can be obtained rather fast. To prove the reliability of the results with only $10^2$ trajectories, we show in Appendix \ref{ap:Nsim} the dependence of our results with the number of trajectories used.

As a last remark, the above analysis could have been performed also in the case in which the motor is instead rotating on average clockwise, i.e. in the synthesis direction. That would require a strong enough external torque to make the motor rotate in reverse, as done in the experimental setup of Ref.~\cite{Hayashi2010_FluctTheoremF1ATPase}. In this situation, Eq.~\eqref{eq:IFT_F1ATPase_StSt} is unchanged, and Eq.~\eqref{eq:IFT_F1ATPase_Gamma} can be easily adapted by settling the reference transition as the synthesis, being now the transition from $i$ to $j$ a transition from $2$ (bound ADP) to $1$ (bound ATP).

\subsection{Bounds to maximal excursions of entropy production} \label{sec:results_bounds}

The expression in Eq. (\ref{eq:WmechInfLowBound}) can be directly adapted to our model for the F1-ATPase motor taking $W_{\mathrm{cycle}}^{\rm c} = \Delta\mu$ and $W_{\mathrm{cycle}}^{\rm m} = W_{\mathrm{ext}}$. We can therefore apply the procedure explained in Sec. \ref{sec:strategies_bounds} to infer the external torque $\tau_{\rm ext}$ through the work performed during a rotation $W_{\rm ext}$, or the temperature $T$. 

\begin{figure}
    \centering
    \includegraphics[width=\linewidth]{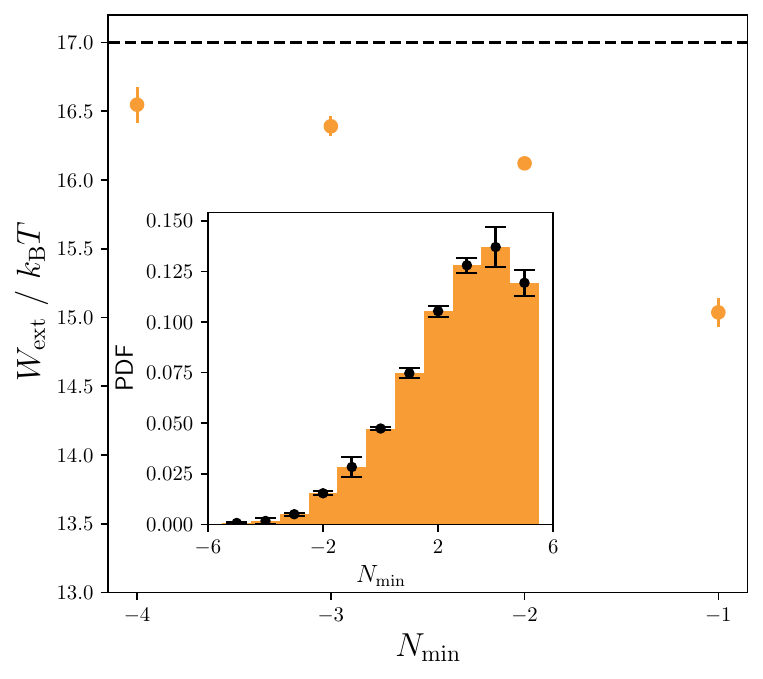}
    \caption{Inference of the external work during a motor rotation using the minimum excursion bound of Eq.~ \eqref{eq:AssBoundWMinus}. The inferred values of $W_{\mathrm{ext}}$ for different values of $N_{\rm min}$, using the corresponding values of  the surpass probability $\alpha$ (yellow circles) are compared with the true value $W= 17~k_B T$ (dashed line). In the inset, we show the histogram of the minima of the extracted work constructed with three independent samples of $10^3$ trajectories, letting the system evolve until a maximum time of $\tau = 0.5~\mathrm{s}$.}
    \label{fig:Results_BoundsWork}
\end{figure}

The results for $W_{\mathrm{ext}}$ are shown in Fig.~\ref{fig:Results_BoundsWork}. Three independent samples of $10^3$ trajectories, letting the system evolve up to a maximum time $\tau = 0.5~\mathrm{s}$, were simulated to construct the histogram of the minimum work performed by the motor (inner panel). The negative tail of the distribution was integrated using the different $N_{\mathrm{min}}$ values showed in the horizontal axis to obtain the probability $\alpha$ that the corresponding bound is reached or surpassed. Using these pairs ($N_{\mathrm{min}}$,~$\alpha$), the external work per step was inferred using Eq.~\eqref{eq:WmechInfLowBound}. 
The best result we obtained comes from $N_{\mathrm{min}}  = -4$, and was $W_{\mathrm{ext}} = 16.6 \pm 0.2~k_{\mathrm{B}}T$. The relative deviation from the true value [see Tab. \ref{tab:exp_values}] is roughly a $2~\%$. We notice that it has lower precision than the results obtained for the IFT at stopping times. The main drawback of this proposal is that obtaining better results requires properly reconstructing
the negative tail of the distribution (see Ref.~\cite{GonzaloEdgar2022_ExtSt}), what would potentially require an unfeasible number of sampled trajectories in a realistic experiment.

In this case, recording the motion of $3\cdot10^3$ trajectories, stopped at $\tau = 0.5~\mathrm{s}$,
would take $25~\mathrm{min}$, which is a rather short time for an experimental measurement. Again, the experimental proposal requires simple and fast measurements. The dependence of the reported results with the number of simulated trajectories used for inference is commented in Appendix~\ref{ap:Nsim}. Similarly to the previous results for the IFT at stopping times in Secs.~\ref{sec:results_IFT_exact} and \ref{sec:results_IFT_approximated}, the inferred values and the accuracy of the method do not change significantly when increasing the number of trajectories by a factor $10$, to $10^3$.

As for the strategies that use the IFT at stopping times, the above analysis could have been shaped for the case in which the motor is biased to rotate on average in the synthesis direction (clockwise). The corresponding expression useful for inference can be obtained from Eq.~(\ref{eq:AssBoundWMinus}) by exchanging 
$W_{\mathrm{ext}}$ and $\Delta\mu$, while the macroscopic efficiency would be in this case $\eta_{\mathrm{syn}} = \Delta\mu/W_{\mathrm{ext}}$. More details about this parallel situation are given in Appendix~\ref{ap:BoundsWork}.

\section{Conclusions} \label{sec:conclusions}

In this work, we have introduced novel strategies for thermodynamic inference in molecular motors based on recent theoretical results of martingale theory for entropy production in stochastic thermodynamics. The proposed strategies apply to generic discrete-state molecular motors transforming chemical work into mechanical work under stationary conditions and acting cyclically over the state space. Here we exemplified and tested the different strategies with calculations and simulations for the relevant case of the F1-ATPase rotatory motor.

The first main strategy we analyzed is based on the IFT for entropy production at stopping times~\cite{Edgar2017_InfStopTimes,Gonzalo2021_StopTGambDemons,Gonzalo2024_AbsIrrev}, which, by observing jumps in a signalling transition, allowed us to infer either the heat exchanged by the motor per cycle of operation, the contribution of absolute irreversibility $\Gamma$, or the temperature of the surrounding medium. As a consequence, such a strategy can be used to obtain the dissipation of the motor, which constrains the sum of all different driving forces acting on the system. 
We thus showed that one can infer individual forces (e.g. an external nonconservative force) when knowing the others (e.g. the chemical potential bias), as in the case of the F1-ATPase rotatory motor working against an external torque. 

The strategy is based on a stopping protocol that allows us to know exactly the entropy production of the stopped trajectories by starting to record the motor motion after a first jump in the signalling transition [see Fig.~\ref{fig:StoppingProtocol}]. A rigorous application of such strategy requires to take into account the presence of absolute irreversibility, which may be often difficult to obtain. Our simulations of the inference strategy for the rotational work performed by the F1-ATPase motor give results compatible with the true value within one standard deviation and with a relative deviation below $0.06\%$. Nevertheless, we also showed that working directly in the NESS, and neglecting both absolute irreversibility and an initial transient, the strategy is simplified while still giving accurate results. Using this approximate method, we obtained an inferred value that is also compatible within one standard deviation with the true value and with a relative deviation below $0.12\%$. {  Importantly, finite statistics estimation of the IFT at stopping times may converge to its theoretical value much faster than the conventional IFT~\cite{Neri19}. Intuitively, this can be made by setting the thresholds to avoid rare trajectories with e.g. large negative entropy production. As a consequence, this strategy becomes suitable for situations with a limited sample of trajectories.}

The second main strategy was instead based on recent bounds to maximal excursions of entropy production in nonequilibrium steady states~\cite{Edgar2017_InfStopTimes,GonzaloEdgar2022_ExtSt}, which also allowed us to infer either the external work performed against an external force acting on the motor, or the temperature of the surrounding medium.
Importantly, the bounds to the entropy production excursions can also be used to bound excursions of other steady-state thermodynamic fluxes, such as the dissipated heat or the work performed against external forces~\cite{GonzaloEdgar2022_ExtSt}. The analytical expressions of the latter and, in particular, of the bound to the minimum work done against an external force allowed us to establish a relation among the external work per step, the net flux in the signalling transition, the probability for the minimum to be reached or surpassed, and the effective temperature of the environment. Sampling trajectories to obtain the probabilities from a frequentist approach, we obtained as a best result for the inference of the external work per step of the F1-ATPase rotatory motor compatible with the true value within two standard deviations and with a relative deviation below $2.4\%$. The main drawback of this second approach is that better results require properly constructing the negative tail of the work minima distribution. Such construction would potentially need an unfeasible number of sampled (recorded) trajectories.

It is worth remarking on the experimental feasibility of the inference strategies proposed here. Given that each jump in the signaling transition of the F1-ATPase motor takes approximately $0.1~\mathrm{s}$ to occur, and that our stopping-time strategies involve at most $N_+ = 30$ jumps, this implies that the experimental samples would need around $3~\mathrm{s}$ in total.
Therefore, it is expected that recording the motor movement along the $10^2$ trajectories takes $300~\mathrm{s} = 5~\mathrm{min}$ on average. In the case of the minimum entropy production strategy, each trajectory was allowed to evolve up to a maximum time $\tau = 0.5~\mathrm{s}$. The $3\cdot10^3$ trajectories needed to reconstruct the minima histogram with reasonable accuracy need around $25~\mathrm{min}$. Therefore, both strategies are based on rather fast real experimental measurements.

{The above numbers for sampling times are comparable to the ones obtained in the experiments using the detailed fluctuation theorem for inference purposes as reported in Ref.~\cite{Hayasi2012}, while the strategies presented here allow more flexibility in the choice of quantities to be inferred and seem to provide relative errors up to 5 times smaller. In this context, it would be also interesting to explore inference methods based on recently derived  detailed fluctuation theorems applicable at stopping times~\cite{Harunari23}.}

As a last remark, we highlight that, while we have focused on the illustrative example of the F1-ATPase rotatory motor, the strategies introduced in this paper are general and can be applied to other molecular motors acting on cycles with at least one signaling transition. In particular, it would be interesting to apply these strategies for the case of myosin (see, for instance, the model proposed for its dynamics in Ref.~\cite{Harunari2022_FewVisibleTransitions}), for discrete-state models of kinesin (see e.g. Refs.~\cite{Liepelt2007,Altaner15,Takaki22}), or to other biochemical processes described as cycles~\cite{Sartori14,Barato14,Bo15,Pinero24}, which may include information flows~\cite{Horowitz14,Ito13,Ito15,Amano22}. Moreover, our inference methods may be applied beyond the case of biomolecular machines, such as for synthetic thermal machines operating in steady-state conditions implemented in nanoelectronic devices~\cite{Cleuren12,Venturelli13,Sothmann15,Erdman18,Sanchez19,Monsel25}. These methods might also be extended to the quantum case by using extensions of the martingale theory for stochastic entropy production for monitored quantum systems~\cite{Manzano19,AVS22}, with possible applications in quantum thermal machines~\cite{Tude24,Almanza24,Hedge24,Aamir25}.

\section*{Acknowledgments}
We wish to thank Léa Bresque, Debraj Das and Édgar Roldán for insighful and fruitful discussions. ANR was funded by the Consejo Superior de Investigaciones Científicas (CSIC) JAE Intro 2023 program. GM acknowledges support from the Ram\'on y Cajal program RYC2021-031121-I funded by MICIU/AEI/10.13039/501100011033 and European Union NextGenerationEU/PRTR, the  CoQuSy project PID2022-140506NB-C21 and 22, and the María de Maeztu project CEX2021-001164-M for Units of Excellence, funded by MICIU/AEI/10.13039/501100011033/FEDER, UE.

\

The supporting data and codes for this article are openly available
from the Ref.~\cite{Data}.

\appendix

\section{Martingale theory in stochastic thermodynamics}
\label{ap:Martingale_Processes}

{  

Martingales are a class of stochastic processes characterized by the property that their expected value at any time, $\tau$, equals its value at a previous time, $t$, when the expectation is conditioned on observations up to that previous time $t$. That is, the process $M(\tau)$ is a martingale if it verifies:
\begin{equation}
    \left< M(\tau) | X_{[0,t]} \right> = M(t),
\end{equation}
with $t\leq \tau$. As a consequence, martingales are unbiased processes that maintain a constant mean over time. Two closely related definitions are submartingales, for which the value of the process at time $t$ is a lower bound of the conditional expectation at $\tau$, $\left< M(\tau) | X_{[0,t]} \right> \geq M(t)$, and supermartingales, for which it is an upper bound, i.e. $\left< M(\tau) | X_{[0,t]} \right> \leq M(t)$.

The introduction of martingales became a breakthrough in probability theory~\cite{Williams}, but they have also played an important role in fields like game theory and quantitative finance. In the last years, the use of martingales in the context of stochastic thermodynamics has begun to be explored~\cite{Edgar2023_MartingalesPhysicists}, mainly motivated by the fact that the exponential of (minus) the total entropy production is a martingale in nonequilibrium stationary processes:
\begin{equation}
    \left< e^{-S_{\mathrm{tot}}(X_{[0,\tau]})} | X_{[0,t]} \right> = e^{-S_{\mathrm{tot}}(X_{[0,t]})}.
    \label{eq:ap_expstot_martingale}
\end{equation}
Indeed, the martingale property above naturally implies the conventional IFT for entropy production at fixed times~\cite{Seifert_2012}. This can be seen by applying the martingale property to the onset of the trajectory, when no entropy could have been produced yet, $S_{\mathrm{tot}}(X_{[0,0]}) = 0$, that is:
\begin{equation}
\begin{aligned}
    \left< e^{-S_{\mathrm{tot}}(X_{[0,\tau]})}\right> &= \sum_{x_0} \left< e^{-S_{\mathrm{tot}}(X_{[0,\tau]})} | x_0 \right> \rho_{x_0}(0) \\
     &= \sum_{x_0} e^{-S_{\mathrm{tot}}(X_{[0,0]})} \rho_{x_0}(0) = 1,
\end{aligned}
\end{equation}
which implies that the martingale is a stronger property of entropy production than the IFT.

Importantly, all results obtained to characterize martingale processes - see, for instance, Doob's seminal work~\cite{book_Doob} - can then be applied to unveil new properties of the statistics of entropy production, which can also be used to further characterize the statistics of other thermodynamic magnitudes of interest. In this context, one useful result of martingales is Doob's optional stopping theorem. It states that the expected value of a martingale at a (bounded) stopping time, $\T$, is equal to its value at the onset of the trajectory, i.e. $\left<M(\T)\right>_\T = \left<M(0)\right>$, where the subscript $\T$ in the bracket emphasizes that the (stopped) trajectories in the average may have different lengths. 

Applying Doob's optional stopping theorem to the martingale $e^{-S_{\rm tot}(X_{0,\tau})}$, allows to obtain the IFT at stopping times for stationary processes, which reads~\cite{Edgar2017_InfStopTimes,Neri19}:
\begin{equation}
    \left<e^{-S_{\mathrm{tot}}(X_{[0,\T]})}\right>_{\T} = 1,
    \label{eq:ap_IFT_stopping}
\end{equation}
which is valid in a NESS for generic stopping conditions for which $\T < \infty$.

The above results can be generalized to arbitrary out-of-equilibrium processes~\cite{Gonzalo2021_StopTGambDemons} by noticing the validity of the more general martingale process:
\begin{equation}
    \left< e^{-S_{\mathrm{tot}}(X_{[0,\tau]}) - \delta_\tau(\tau)} | X_{[0,t]} \right> = e^{-S_{\mathrm{tot}}(X_{[0,t]}) - \delta_\tau(t)},
    \label{eq:ap_expstot_martingale2}
\end{equation}
with the extra term $\delta_\tau(t)$ given in Eq.~\eqref{eq:delta}, which vanishes for $t= \tau$ and for stationary dynamics. Applying Doob's optional stopping theorem to the martingale $e^{-S_{\mathrm{tot}}(X_{[0,t]}) - \delta_\tau(t)}$ gives us the following IFT at stopping times: 
\begin{equation}
    \left<e^{-S_{\mathrm{tot}}(X_{[0,\T]} - \delta_\tau(\T))}\right>_{\T} = 1,
    \label{eq:ap_IFT_stopping}
\end{equation}
which is verified in arbitrary non-equilibrium situations and again for generic stopping conditions for which $\T < \infty$. Moreover, in the presence of absolute irreversibility~\cite{Murashita2014_AbsIrr}, it has been recently shown that Eq.~\eqref{eq:ap_expstot_martingale2} above, becomes a supermartingale~\cite{Gonzalo2024_AbsIrrev}, and the IFT at stopping times with absolute irreversibily becomes: 
\begin{equation}
    \left<e^{-\left[S_{\mathrm{tot}}(\mathcal{T})-\delta(\mathcal{T})\right]}\right>_{\mathcal{T}} = 
    1-\Gamma
    \label{eq:ap_IFT_stopping_Irr}
\end{equation}
with the abolute irreversibility term $0 \leq \Gamma \leq 1$ is given in Eq.~\eqref{eq:Gamma}.

Another important universal result for martingales is Doob's maximal inequality \cite{book_Doob}, which can be used to bound maximal excursions of entropy production in steady state processes. Besides, once the entropy production is bounded, one can also derive bounds for all thermodynamic quantities that are proportional to it.

More precisely, the fact that $e^{-S_{\mathrm{tot}}}$ is a martingale implies that $e^{qS_{\mathrm{tot}}}$ is a submartingale for any real number $q$ verifying $|q| \geq 1$, that is:
\begin{equation}
    \left< e^{qS_{\mathrm{tot}}(X_{[0,t]})} | X_{[0,\tau]} \right> \geq 
    e^{qS_{\mathrm{tot}}(X_{[0,\tau]})}.
\end{equation}
This result can be obtained from Jensen's inequality for conditional expectations, as follows from Ref.~\cite{GonzaloEdgar2022_ExtSt}. Let $M(t)$ be a martingale process. Given that $f(x) = x^{-q}$ is a convex function for $|q| \geq 1$, Jensen's inequality implies that:
\begin{equation}
    \left< f[M(t)] | X_{[0,\tau]} \right> \geq f\left[\left< M(t) | X_{[0,\tau]} \right>\right]
    = f[M(\tau)],
\end{equation}
\noindent what proofs the above inequality if identifying the martingale process as the exponential of minus entropy production, $M(t) = e^{-S_{\mathrm{tot}}(X_{[0,t]})}$.

Doob's maximal inequality states that the probability that the maximum of a non-negative submartingale surpasses a certain value up to a given time is bounded by the mean value of the submartingale
at that time:
\begin{equation}
    P\left(\mathrm{max}_{t\leq\tau} M(t) \geq \lambda\right) \leq 
    \frac{1}{\lambda} \left< M(\tau) \right>,
\label{eq:ap_DoobMaxIneq}
\end{equation}
\noindent with $\lambda > 0$. We have already proven that $e^{qS_{\mathrm{tot}}}$, with  $|q|\geq1$ is a submartingale, and, since it is an exponential,
we also know that it is non-negative. Using Doob's maximal inequality
we can then bound the probability for a maximal excursion 
of the above exponentials of entropy production, and then of the entropy production itself.

To do so, it is convenient to work with two subcases of the above exponentials, namely, $e^{\pm pS_{\mathrm{tot}}}$ with $p\geq 1$. Applying Doob's maximal to these two separate cases:
\begin{equation}
    \begin{gathered}
    P\left(\mathrm{max}_{t\leq\tau} e^{pS_{\mathrm{tot}}(t)} \geq \lambda \right) 
    \leq \frac{1}{\lambda} \left< e^{pS_{\mathrm{tot}}(\tau)} \right>, \\
    P\left(\mathrm{max}_{t\leq\tau} e^{-pS_{\mathrm{tot}}(t)} \geq \lambda \right) 
    \leq \frac{1}{\lambda} \left< e^{-pS_{\mathrm{tot}}(\tau)} \right>,
    \end{gathered}
\end{equation}
\noindent where we simplified the notation by writing $S_{\mathrm{tot}}(X_{[0,t]})$ as 
$S_{\mathrm{tot}}(t)$. 
Since the exponential $e^{\pm S_{\mathrm{tot}}(t)}$ is an injective, 
monotonically increasing ($+p$) or decreasing ($-p$) function of $S_{\mathrm{tot}}(t)$, there is a  one to one correspondence between the value taken by the exponential and the one taken by the entropy production. We can then write $\lambda = e^{\pm ps}$, with $s$ a threshold for $S_{\mathrm{tot}}$. In the case $e^{+pS_{\mathrm{tot}}}$, that functional increases as the entropy production increases, and the maximal excursion of the exponential implies a maximal excursion of the entropy production, which enables us to write:
\begin{equation}
    P\left(\mathrm{max}_{t\leq\tau} S_{\mathrm{tot}}(t) \geq s \right) \leq 
    e^{-ps} \left< e^{pS_{\mathrm{tot}}(\tau)} \right>.
    \label{eq:ap_PExtUpperStot}
\end{equation}
On the other hand, in the case $e^{-pS_{\mathrm{tot}}}$, it increases as the entropy production decreases, and hence the probability of the exponential reaching some maximal value is related to the probability of the entropy production reaching some minimum value. Equivalently to the case $+p$, we can write for this case:
\begin{equation}
    P\left(\mathrm{min}_{t\leq\tau} S_{\mathrm{tot}}(t) \leq -s \right) \leq
     e^{-ps} \left< e^{-pS_{\mathrm{tot}}(\tau)} \right>.
    \label{eq:ap_PExtLowerStot}
\end{equation}
These results provide the basis to our strategy based on maximal excursions of entropy production in Sec.~\ref{sec:results_bounds}. They can also be used to find explicit expressions for optimal upper and lower thresholds that are not reached with a certain confidence level~\cite{GonzaloEdgar2022_ExtSt}.
}

\section{Bounds for the output mechanical work} \label{ap:BoundsWork}

In this Appendix we adapt the bounds to the survival probability in Eq.~(\ref{eq:bounds_SurvStot}) of Ref.~\cite{GonzaloEdgar2022_ExtSt} to our model. The aim is to obtain bounds to the maximal excursions of the output mechanical work performed by the molecular motor against an external non-conservative force.

In order to set a correspondence between the maximal excursions of the entropy production and those of the output mechanical work, we need to write the entropy production in terms of the mechanical work. In Eq.~(\ref{eq:StotExplicit}) we wrote the entropy production in terms of the heat exchanged with the reservoirs, which, upon using the first law at the trajectory level, allows us to relate heat and work throughout a single trajectory:
\begin{equation}
    \begin{gathered}
    S_{\mathrm{tot}}(t) = \Delta S_{\rm sys}(t) - \beta \left[\Delta E (t) + W^{\mathrm{c}}(t) - W^{\mathrm{m}}(t)\right] \\
    = \beta\left[W^{\mathrm{m}}(t) - W^{\mathrm{c}}(t) - \Delta F (t)\right],
    \end{gathered}
\end{equation}
where we split the work into the contribution coming from output mechanical driving $W^{\mathrm{m}}(t)$, and that coming from the input chemical one $W^{\mathrm{c}}(t)$.
In the last equation, we have identified a stochastic entropy of the system, $S_{\rm sys}(t)$, that accounts for the surprisal and the mesoscopic entropy of the system mesostates:
\begin{equation}
    S(t) = - \ln \rho_{x_t}(t) + S^{\mathrm{int}}_{x_t}(t),
\end{equation}
where, as commented before, the latter is assumed to be zero without loss of generality. Finally we  have also introduced the stochastic free energy, $F(t) = E(t)-k_{\mathrm{B}}T S_{\rm sys}(t)$.

In the stationary state, all non-extensive quantities in time such as energy and entropy changes vanish on average, and therefore, all time-extensive currents become on average proportional to each other. In particular, the average input chemical work and the average output mechanical work, are related by the macroscopic efficiency of the motor, $\eta$:
\begin{equation}
    \begin{gathered}
    \eta = \frac{\left< W^{\mathrm{m}}(t) \right>}{\left< W^{\mathrm{c}}(t) \right>} 
    = \frac{W_{\mathrm{ext}}}{\Delta\mu}.
    \end{gathered}
    \label{eq:efficiency}
\end{equation}
Note that the macroscopic efficiency is time-independent. We can then write 
$\left< W^{\mathrm{c}}(t) \right> = \eta^{-1} \left< W^{\mathrm{m}}(t) \right>$, and express the chemical work in a trajectory up to a time $t$ by:
\begin{equation}
    W^{\mathrm{c}}(t) = \eta^{-1} W^{\mathrm{m}}(t) + \xi(t),
    \label{eq:WChem}
\end{equation}
\noindent being $\xi(t)$ a non-extensive in time stochastic variable with zero mean, $\left<\xi(t)\right> = 0$. Going back to the total entropy production, using Eq.~\eqref{eq:WChem}, we can now rewrite it as:
\begin{equation*}
    S_{\mathrm{tot}}(t) = \beta \left[(1 - \eta^{-1})W^{\mathrm{m}}(t) - \Delta F (t) +\xi(t)\right]
\end{equation*}
\noindent which allows us to relate the global maximum and minimum of entropy production and mechanical work as:
\begin{equation}
    \begin{gathered}
        \mathrm{max}_{t\leq\tau} S_{\mathrm{tot}}(t) \leq \beta\left(\frac{1-\eta}{\eta}\right) 
        \mathrm{max}_{t\leq\tau} W^{\mathrm{m}}(t) + \kappa, \\
        \mathrm{min}_{t\leq\tau} S_{\mathrm{tot}}(t) \geq \beta\left(\frac{1-\eta}{\eta}\right)
        \mathrm{min}_{t\leq\tau}  W^{\mathrm{m}}(t) - \kappa,
    \end{gathered}
\end{equation}
\noindent where $\kappa = \mathrm{max}_{t\leq\tau} \left|\beta \left[\Delta F (t)-\xi(t)\right]\right| \geq 0$ is the maximal excursion in modulus of the non-extensive part of the entropy production. Recall that the actual output of the motor $W^{\mathrm{m}}(t)$, is positive when working in the hydrolysis direction.

Once the correspondence between maximal excursions has been found, one can apply the results for 
the entropy production in Eq.~(\ref{eq:bounds_SurvStot}) by changing variables. The particular characterization of extreme statistics in which we are interested is the bound to the output mechanical work that will only be exceeded with a certain probability. Sudden strong fluctuations could, for instance, damage the motor, so it is important to ensure that the experimental setup is safe enough to make reliable measurements. We already know the probability for the entropy production to surpass upper and lower thresholds from Eq.~(\ref{eq:bounds_SurvStot}). The survival probability of the entropy production not surpassing them is simply one minus that probability. To that probability we can impose a certain value, say $1-\alpha$, where $\alpha$ is the probability we allow the motor to surpass the threshold:
\begin{equation}
    \begin{split}
        &P\left[W^{\rm m}_\mathrm{max}(t) \leq w_+\right] = 1-P\left[W^{\rm m}_{\max}(t) \geq w_+\right] \\
        &= 1-P\left[\mathrm{max}_{t\leq\tau} S_{\mathrm{tot}}(t) \geq s_+\right] = 1-\alpha, \\
    \end{split}
\end{equation}
with upper threshold equivalence
\begin{eqnarray}
    s_+ = \beta\left(\frac{1-\eta}{\eta}\right)w_+ + \kappa.
\end{eqnarray}
and where we denoted $W^{\rm m}_\mathrm{max}(t) = \max_{t\leq \tau} W^{\rm m}(t)$ the maximum of $W(t)$ within the interval $[0,\tau]$. Equivalently, for the lower bound we have:
\begin{equation}
    \begin{split} 
&P\left[W_{\min}^{\mathrm{m}}(t) \geq -w_-\right] = 1-P\left[W_{\min}^{\mathrm{m}}(t) \leq -w_-\right] \\
&= 1-P\left[\mathrm{min}_{t\leq\tau} S_{\mathrm{tot}}(t) \leq -s_-\right] = 1-\alpha,
    \end{split}
\end{equation}
with lower threshold
\begin{eqnarray}
    s_- = \beta\left(\frac{1-\eta}{\eta}\right)w_- - \kappa,
\end{eqnarray}
and where we analogously denoted $W_{\min}^{\mathrm{m}}(t) = \mathrm{min}_{t\leq\tau} W ^{\mathrm{m}}(t)$ the minimum of the mechanical work during the interval $[0,\tau]$.

Using the expressions for the probabilities of the entropy production in Eq.~(\ref{eq:bounds_SurvStot}) with the corresponding values of $s$ in each case, the equation for the upper, $w_+$, and the lower bounds, $w_-$ are:
\begin{equation}
    \begin{gathered}
    1-e^{-p\left[\beta\left(\frac{1-\eta}{\eta}w_+ + \kappa\right)\right]}\left<e^{p S_{\mathrm{tot}}(t)}\right> = 1-\alpha,\\
    1-e^{-p\left[\beta\left(\frac{1-\eta}{\eta}w_- - \kappa\right)\right]}\left<e^{-p S_{\mathrm{tot}}(t)}\right> = 1-\alpha,
    \end{gathered}
\end{equation}

\noindent which, after some basic algebra, give the following result for the optimal famility of thresholds not surpassed during the interval $[0,\tau]$ with probability $1-\alpha$:
\begin{equation}
    \begin{split}
    w_{\pm}^{(p)} &= \frac{T\eta}{1-\eta}\left[\ln\frac{\left<e^{\pm p S_{\mathrm{tot}}(t)}\right>^{1/p}}
    {\alpha^{1/p}}\mp \kappa\right] \\
    &\approx \frac{T\eta}{1-\eta}\ln\frac{\left<e^{\pm p S_{\mathrm{tot}}(t)}\right>^{1/p}}
    {\alpha^{1/p}},
    \label{eq:NonAproxBoundsWork}
    \end{split}
\end{equation}

\noindent where the last approximated equality follows whenever $\kappa$ can be neglected. Because of the submartingale property, the moment of order $p$ of the exponential of the entropy production is always increasing in time, unlike $\kappa$, a non-extensive in time quantity. Hence, for sufficiently large times, the term $\kappa$ will be negligible. Moreover, simply using smaller values of $\alpha$ will also lead to $\kappa$ negligible. Note that the last equation is actually a family of bounds, one for each value of $p$. The final result is actually the tightest bound of the family, formally given by:
\begin{equation}
    \begin{gathered}
    w_{\pm} = \frac{T\eta}{1-\eta}\underset{p\geq1}{\mathrm{min}}\ln\frac{\left<e^{\pm p S_{\mathrm{tot}}(t)}\right>^{1/p}}
    {\alpha^{1/p}} \\
    = \frac{T\eta}{1-\eta}\ln\left[\underset{p\geq1}{\mathrm{min}}\frac{\left<e^{\pm p S_{\mathrm{tot}}(t)}\right>^{1/p}}
    {\alpha^{1/p}}\right].
    \end{gathered}
    \label{eq:BoundsWork_ap}
\end{equation}
Although theoretically useful, the above bounds are not experimentally accessible. They generically need the mean value of the exponential of the entropy production and other moments, which requires full knowledge of the entropy production in the system. If such information was available, no inference would be needed. Whether these bounds become useful for our purposes depends on the possibility of simplifying them in some regime such that not that much information is required to obtain them. In Ref. \cite{GonzaloEdgar2022_ExtSt}, 
this simplification was achieved in the long-time limit, as follows from large deviation theory \cite{Touchette2009_LDT}:
\begin{equation}
    \begin{gathered}
     t \lim_{t\to\infty} \frac{1}{t} 
    \ln\left<e^{\pm p S_{\mathrm{tot}}(t)}\right> = t \lambda(\pm p),
    \end{gathered}
\end{equation}

\noindent where $\lambda(\pm p)$ is the scaled cumulant generating function of the exponential of minus the entropy production.
The minimization with respect to $p$ was already solved in Ref. \cite{GonzaloEdgar2022_ExtSt} using the convexity of the
scaled cumulant generating function and the relation $\lambda(-p) = \lambda(p+1)$, yielding as optimal thresholds:
\begin{equation}
    w_{\pm} = \frac{T\eta}{1-\eta}\left[t\lambda(\pm 1)-\ln\alpha\right].
    \label{eq:AssymptoticBounds}
\end{equation}
Notice that the optimal upper bound still needs a reconstruction of the mean of the exponential of the entropy production through $\lambda(1)$. However, the optimal lower bound is achieved for $p=-1$, and we have $\lambda(-1) = 1$ from the standard IFT~\cite{Seifert2005_EntropyProd_IFT}.
Therefore, in this case the bound saturates to a time-independent value:
\begin{equation}
    w_{-} = -\frac{T\eta}{1-\eta}\ln\alpha,
    \label{eq:AssBoundWMinus_ap}
\end{equation}
akin to the infimum law results for entropy production derived in Ref.~\cite{Edgar2017_InfStopTimes}.

Finally, if the motor was working in the synthesis direction, the expressions for the optimal bounds of the extracted mechanical work, Eqs.~(\ref{eq:AssymptoticBounds}) and (\ref{eq:AssBoundWMinus_ap}), would hold for the chemical work (actually for minus the input chemical work). The only difference would be the expression of the macroscopic efficiency, which in that case would be inverted:
\begin{equation}
    \eta_{\rm syn} = \frac{\left< W^{\mathrm{c}}(t) \right> }{\left< W^{\mathrm{m}}(t) \right>} = 
    \frac{\Delta\mu}{W_{\mathrm{ext}}}.
\end{equation}

\section{Experimental values for the parameters} \label{ap:ExpValues}

In this appendix, we justify the values of the parameters used in our calculations and simulations as shown in Tab.~\ref{tab:exp_values}. We used the energetic characterization of Ref.~ \cite{Oster1998_EnergyTransductionF1ATPase} and
Ref. \cite{Toyabe2010_Energetics}, as well as the time scales of the motor from Ref. ~\cite{Noji1997_DirectObsF1ATPase}. 

Because typically only the hydrolysis and synthesis are observable due to the associated macroscopic rotation of the motor, the values of the parameters for the remaining transitions are adapted to be consistent with those for the observable ones. To begin with, no information was found about the intrinsic entropy of each motor state.  
Therefore, as an approximation, all states were considered isoentropic, $\Delta S^{\mathrm{int}}_{ij} = 0$, and the energy changes to be found are then internal energy changes. 

In Refs.~\cite{Oster1998_EnergyTransductionF1ATPase,Toyabe2010_Energetics} an efficiency close to 100\% for the F1-ATPase was obtained. This efficiency was defined as the ratio of heat exchanged in a hydrolysis transition to the energy used by the motor to rotate, provided by the chemical potential difference between a molecule of ATP and a molecule of ADP, $\Delta\mu = \mu_{\mathrm{ATP}}-\mu_{\mathrm{ADP}}$. 
In Ref.~\cite{Toyabe2010_Energetics}, a compilation of previous measurements of $\Delta\mu$ was presented together with their mean, roughly $\Delta \mu = 75~\mathrm{pN~nm}$, or, rounding up, $18~k_{\mathrm{B}}T$, with an external temperature $T = 25^{\circ} \mathrm{C}$. The error bars of the experiments made the measured heat compatible with $\Delta\mu$. However, in some cases, the measured heat was even higher than $\Delta\mu$. That effect was due to experimental errors, since the motor cannot dissipate more energy than what it has available. In order to obtain an efficiency close to 100\% while being consistent with the motor functioning, we set $\Delta E_{x_1 x_2} = -17.5~k_{\mathrm{B}}T$, with $\Delta E_{x_1 x_2} = -Q^{\mathrm{diss}}_{x_1 x_2}$.

The remaining values of the energy differences for the transitions were chosen based on the previous one. Importantly, according to several sources (as e.g. Ref.~\cite{Toyabe2011}) we must ensure that, under normal conditions, bounding an ATP is the rate-limiting step. That is why we set the energy difference between the empty and the ATP bounded state to be the largest energy gap, $\Delta E_{x_0 x_1} = 20~k_{\mathrm{B}}T$. The energy difference between the empty and the ADP bounded state is then chosen by consistency with the other values, ensuring
that the total internal energy change in one cycle is zero.

The external work performed by the motor against the external torque does not aim to model any particular intrinsic property of the F1-ATPase motor, but depends on its external manipulation. Therefore, we take it as a free parameter. The value of $17~k_{\mathrm{B}}T$ was chosen to be of the same order as the energy differences, so as to be relevant in the dynamics. We took it to not exceed $\Delta\mu$, which would make the motor rotate in reverse. As already mentioned
in the introduction, the motor can rotate either hydrolysing or synthetising ATP. The study of both modes of operation is interesting, but the analysis we performed in this work is equivalent for both cases, so we only focus on one of them, arbitrary, the hydrolysis mode.

Finally, the only available measurement of the time scale of the motor dynamics is its mean rotational speed. However, a full 360º rotation includes three net hydrolysis or synthesis steps in our model, which implies three cycles in either direction once. Given the lack of information about the rates of individual transitions, the first approximation was to consider $k_{\mathrm{ATP}} = k_{\mathrm{ADP}} = k_{HS} = \gamma$, with $\gamma$ the parameter that establishes the time scale of our simulations. Under normal conditions, the motor is expected to work in its stationary state, so we need the rate at which cycles are completed in the stationary regime and 
relate it to the experimental measurements. We obtained in Appendix \ref{ap:MasterEq} that the rate of cycles is given by Eq.~\eqref{eq:stflux}. According to the 
first experimental characterization of the F1-ATPase in Ref.~\cite{Noji1997_DirectObsF1ATPase}, in the absence of external forces ($W_{\mathrm{ext}} = 0$) the motor rotates in the hydrolysis direction doing around $K_{\rm revs} = 17$ r.p.s. We hence approximate the value of the rate $\gamma$ as:
\begin{eqnarray}
    &K_{x_1 x_2}\pi_{x_1}-K_{x_2 x_1}\pi_{x_2} = \nonumber \\ 
    &\gamma\left( \pi_{x_1} e^{-\frac{\beta}{2}\Delta E_{x_1 x_2}}
    - \pi_{x_2} e^{\frac{\beta}{2}\Delta E_{x_1 x_2}}\right) = 
    K_{\rm revs}\cdot 3~\frac{\mathrm{cycles}}{\mathrm{rev}}, \nonumber \\
 &\gamma = \frac{3K_{\rm revs}}{\pi_{x_1} e^{-\frac{\beta}{2}\Delta E_{x_0 x_1}}
    - \pi_{x_2} e^{\frac{\beta}{2}\Delta E_{x_0 x_1}}} \approx 165~\mathrm{Hz}.
\end{eqnarray}

\section{Master equation and steady state} \label{ap:MasterEq}

In this Appendix we provide details on the Markovian dynamics of the F1-ATPase motor model and the 
calculations for obtaining the stationary probability distribution and related quantities.
The master equation for the system, in matrix form, is given by: 
\begin{equation}
    \begin{gathered}
        \deriv{}{t}\begin{pmatrix} \rho_{x_0}(t) \\ \rho_{x_1}(t) \\ \rho_{x_2}(t) \end{pmatrix} = 
        \mathbb{K}  
         \begin{pmatrix} \rho_{x_0}(t) \\ \rho_{x_1}(t) \\ \rho_{x_2}(t) \end{pmatrix},
        \label{eq:MastEq_Matrix}
\end{gathered}
\end{equation}
where $\{\rho_{x_0}(t), \rho_{x_1}(t), \rho_{x_2}(t)\}$ are the time-dependent probabilities of states $x_0$, $x_1$ and $x_2$, respectively, and the rate matrix reads:
\begin{widetext}
\begin{equation}
        \mathbb{K} = 
        \begin{pmatrix}
            -(K_{x_0 x_1} \!+\! K_{x_0 x_2}) \!\!&\!\! K_{x_1 x_0} \!\!&\!\! K_{x_2 x_0} \\
            K_{x_0 x_1} \!\!&\!\! -(K_{x_1 x_0} \!+\! K_{x_1 x_2}) \!\!&\!\! K_{x_2 x_1} \\
            K_{x_0 x_2} \!\!&\!\! K_{x_1 x_2} \!\!&\!\! -(K_{x_2 x_0} \!+\! K_{x_2 x_1})
        \end{pmatrix}.
 \label{eq:ratematrix}
\end{equation}
\end{widetext}
After setting the left-hand side of Eq.~\eqref{eq:MastEq_Matrix} to zero, the stationary distribution can be found by solving the eigenvector associated to the null eigenvalue of the matrix. The result is given by:
\begin{equation}
    \begin{split}
        N\pi_{x_0} = K_{x_2 x_0} K_{x_1 x_2} + K_{x_1 x_0} (K_{x_2 x_0} + K_{x_2 x_1}), \\
        N\pi_{x_1} = K_{x_2 x_0} K_{x_0 x_1} + K_{x_2 x_1} (K_{x_0 x_1} + K_{x_0 x_2}), \\
        N\pi_{x_2} =  K_{x_1 x_0} K_{x_0 x_2} + K_{x_1 x_2}(K_{x_0 x_1} + K_{x_0 x_2}), \\
    \end{split}
    \label{eq:St_Prob}
\end{equation}
with normalization parameter $N :=  K_{x_1 x_0} (K_{x_2 x_0} + K_{x_2 x_1} + K_{x_0 x_2}) + (K_{x_0 x_1} + K_{x_0 x_2}) (K_{x_2 x_1} + K_{x_1 x_2}) + K_{x_2 x_0} (K_{x_0 x_1} + K_{x_1 x_2})$.

We remark that the stationary probabilities $\{\pi_{x_0}, \pi_{x_1}, \pi_{x_2} \}$ calculated above are often not accessible from the observation of the hydrolysis and synthesis transition. However, in Appendix~\ref{ap:ExpValues} we show a method to adapt the information from the observable transitions to infer information of the other transitions, obtaining a reliable description of the motor dynamics. Those are also the results we used alongside the theoretical rates in Eq.~(\ref{eq:rates}) and the experimental values provided in Tab.~\ref{tab:exp_values} to simulate the motor dynamics.

An important result which derives from the above description is the time scale at which the system completes a cycle in the stationary state, since it gives the timescale of the motor dynamics. It is given by the probability fluxes in the stationary regime, which become equal for all transitions. Choosing any pair of equations in the system of Eq.~(\ref{eq:MastEq_Matrix}) and setting the l.h.s. to zero, we obtain: 
\begin{equation}
\begin{split}
K_{x_0 x_1}\pi_{x_0}-K_{x_1 x_0}\pi_{x_1} = K_{x_1 x_2}\pi_{x_1}-K_{x_2 x_1}\pi_{x_2} \\ = K_{x_2 x_0}\pi_{x_2}-K_{x_0 x_2}\pi_{x_0}.
\end{split} \label{eq:stflux}
\end{equation}
We use the expression above in Appendix~\ref{ap:ExpValues} to set the time scale of the motor dynamics in the absence of external torque, leading to $W_{\mathrm{ext}} = 0$. (We notice that in all figures we instead have $W_{\mathrm{ext}} = 17~k_{\mathrm{B}}T$.) 

\begin{figure}
    \centering
    \includegraphics[width = \linewidth]{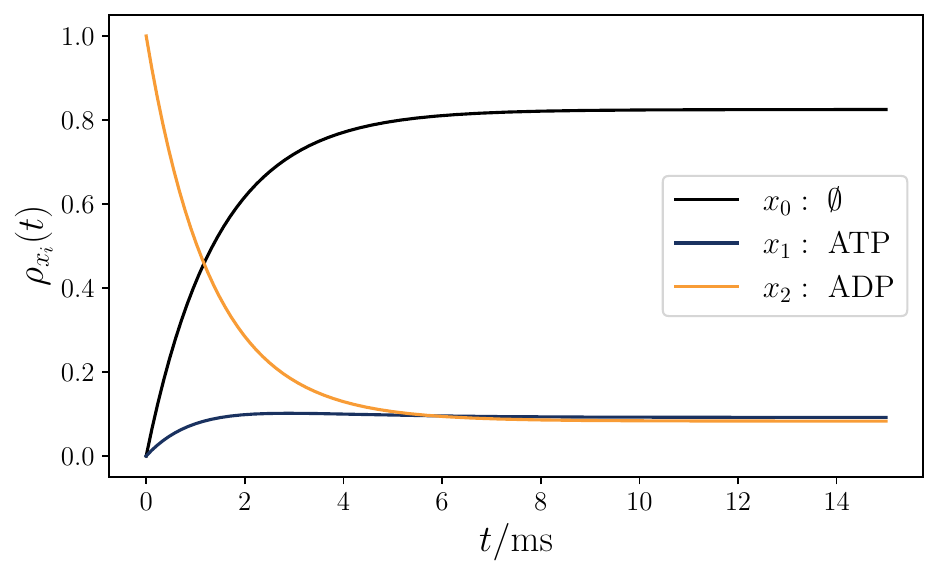}
    \caption{Time evolution of the probabilities of each state of the motor starting from the initial distribution: $\rho_{x_0}(0) = \rho_{x_1}(0) = 0$, and $\rho_{x_2}(0) = 1$. The time-scales of relaxation towards the stationary values $\pi_{x_0}$, $\pi_{x_1}$ and $\pi_{x_2}$ as given in Eq.~\eqref{eq:St_Prob} are of the order of $10~\mathrm{ms}$.}
    \label{fig:Probabilities}
\end{figure}

The general expression for the full-time evolution of the probabilities for arbitrary initial conditions is not actually useful. Instead, we are interested in obtaining an order of magnitude of the time at which the system reaches the stationary state. We focus on the initial conditions $\rho_{x_0}(0) = \rho_{x_1}(0) = 0$, and $\rho_{x_2}(0) = 1$, that is, when the system begins the evolution in the bound ADP state, as in the stopping protocol of Fig.~\ref{fig:StoppingProtocol}. The time evolution of the probabilities is displayed in Fig.~\ref{fig:Probabilities}. The master equation was solved using the \textit{solve()} function from the \textit{DifferentialEquations.jl} package in Julia. It used a composite method combining \textit{Tsit5()} and \textit{Rosenbrock23()} methods.

The system evolves until it reaches a stationary state in which the empty state accumulates
around the 80\% of the probability. Rounding up, the result for the stationary distribution
is: $\pi_{x_0} = 0.825$, $\pi_{x_1} = 0.092$ and $\pi_{x_2} = 0.083$. This is consistent with the picture in which the rate for bounding of ATP is the limiting factor: the system waits in the empty state the majority of the time, until one ATP is attached to the motor, and then the system quickly rotates hydrolysing it, until the empty state is reached again.

The timescale at which the system completes a cycle for non-zero external torque, with $W_{\mathrm{ext}} = 17~k_{\mathrm{B}} T$ can be obtained from Eq.~(\ref{eq:St_Prob}), and is given by $K_{x_0 x_1}\pi_{x_0}-K_{x_1 x_0}\pi_{x_1} \simeq 8.8~\mathrm{s^{-1}}$. 
That is, the motor completes a cycle, on average, in a time of the order of $0.1~\mathrm{s}$.
As shown in Fig.~\ref{fig:Probabilities}, the relaxation to the stationary regime takes less than $10~\mathrm{ms} = 0.01~\mathrm{s}$, which is already an order of magnitude smaller than the time scales of the cycles. This difference between timescales implies that the observable behaviour of the motor (i.e. the rotational steps of the motor) always occurs in the stationary regime.

\section{Dependence of the results with the number of trajectories} \label{ap:Nsim}

\begin{figure}[b]
    \centering
    \includegraphics[width = \linewidth]{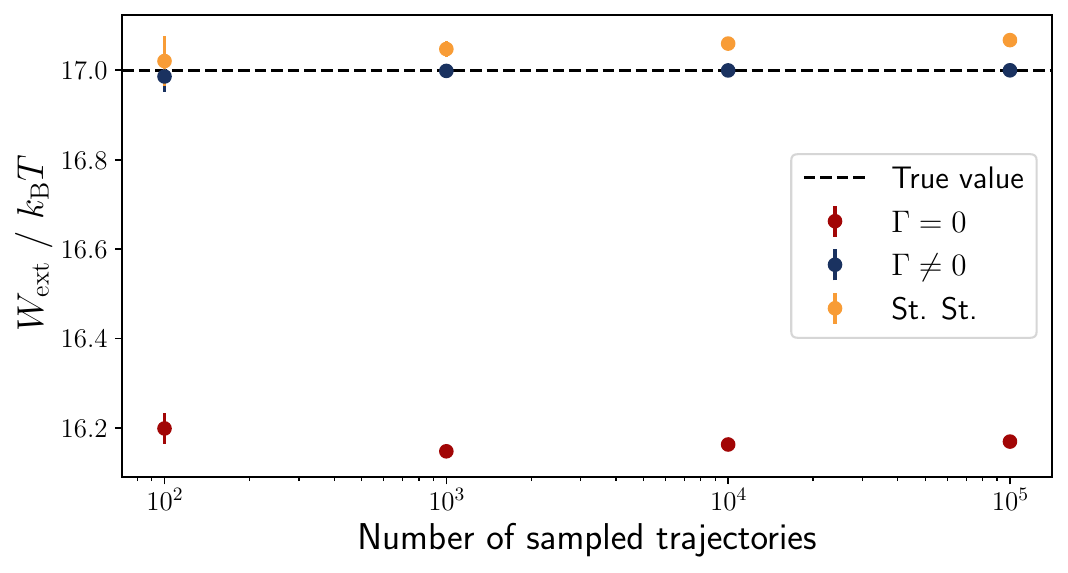}
    \caption{Dependence of different inference results for $W_{\rm ext}$ in the F1-ATPase rotatory motor based on the IFT at stopping times with the number of sampled trajectories in the simulations. Blue dots corresponds to results using the full IFT [Eq.~\eqref{eq:IFT_F1ATPase_Gamma}], red dots correspond to the case where the absolute irreversibility is neglected [Eq.~\eqref{eq:IFT_F1ATPase_Gamma}] and yellow dots to the approximated expression in Eq.~\eqref{eq:IFT_F1ATPase_StSt} neglecting the transient dynamics.}
    \label{fig:WextIFT_Nsim}
\end{figure}

In this appendix, we explicitly show the dependence of the results obtained in the main text with the number of trajectories used in the simulations to show that, even though statistical errors decrease when increasing the number of trajectories, the change in the final result is negligible.

The inference results for the output work obtained with the strategy based on the IFT at stopping times is displayed in Fig.~\ref{fig:WextIFT_Nsim} as a function of the number of trajectories. There we can observe that when the number of trajectories reaches $10^3$, the results already become quite stable when comparing with the values generated with $10^4$ and $10^5$ trajectories. Indeed all the results do not differ significantly from those obtained with the smaller number of ($10^2$) trajectories.

Similarly, the inference results using the bounds to the survival probability of the entropy production display are robust. The only significant difference in the results when increasing the number of sampled trajectories, is a change in the shape of the work histogram that appears from $10^2$ to $10^3$ sampled trajectories. Except for that detail, the results are quite stable and do not differ significantly from those obtained with $10^2$ trajectories.

\bibliography{bibliography.bib}

\end{document}